\documentclass[usenatbib]{mnras}
\usepackage{natbib,amssymb,amsmath,times,graphicx,url,aas_macros}
\usepackage{setspace}      
\usepackage{epstopdf}       
\usepackage{verbatim}       
\usepackage{bm}		  
\usepackage{fleqn}		  
\usepackage{enumitem}
\usepackage[normalem]{ulem} 
\usepackage{Wurster_macros}  

\defcitealias{WursterBatePrice2019}{WBP2019}
\defcitealias{WursterRowan2023a}{Paper I}
\newcommand{\tfinal}{$t_\text{final}$}

\newcommand{\ifive}{\textit{I05}}

\newcommand{\itwenty}{\textit{I20}}

\newcommand{\nfive}{\textit{N05}}

\newcommand{\ntwenty}{\textit{N20}}
\newcommand{\hyd}{\textit{Hyd}}

\newcommand\cnfive[1]{\textit{N05}c\textit{#1}}
\newcommand\snfive[1]{\textit{N05 [{#1}]}}
\newcommand\cifive[1]{\textit{I05}c\textit{#1}}
\newcommand\sifive[1]{\textit{I05 [{#1}]}}

\title[Star-forming environments II]{Star-forming environments in smoothed particle magnetohydrodynamics simulations II: Re-simulating isolated clumps to determine equivalence of extracted clumps and parent simulations}

\author[Wurster \& Rowan]{James Wurster$^{1}$\thanks{james.wurster.astro@gmail.com} and Connar Rowan$^{2,1}$\\
$^{1}$Scottish Universities Physics Alliance (SUPA), School of Physics and Astronomy, University of St. Andrews, North Haugh, St Andrews, Fife KY16 9SS, UK \\
$^{2}$Rudolf Peierls Centre for Theoretical Physics, Clarendon Laboratory, Parks Road, Oxford, OX1 3PU, UK \\
}
\date{Submitted: Revised: Accepted: }
\pagerange{\pageref{firstpage}--\pageref{lastpage}} \pubyear{2023}

\begin{document}
\label{firstpage}
\bibliographystyle{mnras}
\maketitle

\begin{abstract}
What is the numerical reproducibility of a stellar system (including its discs) when evolving only a sub-set of (partially-evolved) smoothed particle hydrodynamics (SPH) particles?  To investigate this, we modelled the evolution of 29 star forming clumps that were extracted from our previous simulations that investigated the formation and early evolution of low-mass star clusters.  These clumps were evolved using a three-dimensional smoothed particle radiation magnetohydrodynamics code, where we included or excluded non-ideal magnetohydrodynamics to match the cluster simulation.  While star formation proceeded as expected, we were unable to identically reproduce any of the systems present at the end of the cluster simulations.  However, the final distributions of stellar mass, stellar system mass, disc mass, and disc radii were reproduced statistically; unfortunately, the distribution of average magnetic field strengths in the discs was not reproduced statistically, but this may be a result of our updated algorithms governing the evolution of the magnetic field.  Therefore, given that our clumps yield stellar masses that are statistically similar to those in the original low-mass star clusters, we have demonstrated that we can statistically reproduce systems (aside from their magnetic field strength) by evolving a subset of SPH particles.  Therefore, clumps such as these can be used as initial conditions to investigate the formation of isolated stars from less-contrived initial environments.
\end{abstract}


\section{Introduction}
\label{intro}
High-resolution numerical simulations of isolated star formation are very useful to investigate the star formation process and the associated physical processes.  These simulations are often initialised using idealised initial conditions to minimise degeneracies when adding new physics.  Although modelling star formation in these pristine initial environments permits a detailed investigation into the new physical process, parameters, or initial conditions, it carries the risk of incorrect conclusions resulting from the pristine environment itself.  For example, the protostellar disc formation community grappled with the magnetic braking catastrophe \citepeg{AllenLiShu2003,MellonLi2008} for over a decade until \citet{Tsukamoto+2015hall} and \citet{WursterPriceBate2016,Wurster+2022} robustly showed that the Hall effect was required to prevent the catastrophe.  However, when modelling low-mass star clusters, where the star-forming clumps were influenced by their  larger-scale environment, it was discovered that the magnetic braking catastrophe never even existed \citepeg{Seifried+2013,WursterBatePrice2019}!   This demonstrated the importance of using a realistic initial environment.  

Given the high resolution permitted when modelling isolated star formation, these studies must continue.  However, great care must be taken when setting the initial environment.  
There have been many studies that investigated various initial (free) parameters, including
the amount and type of kinetic energy, namely turbulence \citepeg{Seifried+2012,Seifried+2013,Joos+2013,TsukamotoMachida2013,MatsumotoMachidaInutsuka2017,WursterLewis2020d,WursterLewis2020sc} and rotation \citepeg{WursterLewis2020d,WursterLewis2020sc},
the internal energy \citepeg{Tsukamoto+2018}, 
the magnetic field strength \citepeg{BateTriccoPrice2014,WursterPriceBate2016,MachidaHiguchiOkuzumi2018}, 
the magnetic field geometry, including fields misaligned with the rotation axis \citepeg{Machida+2006,JoosHennebelleCiardi2012,LewisBatePrice2015,Tsukamoto+2017,Tsukamoto+2018,HiranoMachida2019,Hirano+2020} and turbulent magnetic fields \citepeg{GerrardFederrathKuruwita2019}, 
the comic ray ionisation rate \citepeg{WursterBatePrice2018sd,WursterBatePrice2018ion,KobayashiTakaishiTsukamoto2023}, and
the inclusion/exclusion of the various non-ideal magnetohydrodynamics processes \citepeg{TomidaOkuzumiMachida2015,Tsukamoto+2015oa,Tsukamoto+2015hall,WursterPriceBate2016,WursterBateBonnell2021}.  
By varying the parameter under investigation, these studies produced a variety of systems with a variety of environmental characteristics.  While these studies provided great insights into the effect of these parameters, for each process, what is the optimal and most realistic parameter?  How should these parameters be selected to produce a realistic star forming core?

To mitigate these questions, higher-mass simulations can be performed \citepeg{Bate2012,Seifried+2013,WursterBatePrice2019}, where stars and discs form and evolve in a clustered environment.  In addition to providing more realistic details about star and disc formation and evolution given their less-contrived environment\footnote{i.e., while the large-scale simulation itself is initialised by prescribed (and possibly idealised) initial conditions, star formation itself starts from already evolved gas, hence our label of `less-contrived.'}, this may provide details on the star formation mechanism itself, namely whether it occurs via core collapse \citep{MckeeTan2002,MckeeTan2003}, competitive accretion \citep{Zinnecker1982,Bonnell+2001ca,Bonnell+2001acc}, or inertial-inflow \citep{Padoan+2020}.  The latter two models suggest that gas from well-beyond the initial star-forming core ultimately joins and influences the system; this indicates that modelling star formation in isolated cores represents an incomplete environment, even if selecting the most reasonable parameters.  Although this provides evidence for the requirement of modelling higher-mass systems, modelling higher-mass systems as in the above references necessarily means doing so at lower numerical resolution.   While the evolution of stars and their discs will be more realistic and gas will be permitted to flow in from great distances, many features will be under-resolved or not resolved at all \citep{WursterBateBonnell2021}.  Nonetheless, this is a necessary compromise.  

As an alternative compromise to increasing mass and lowering resolution, it should be possible to extract regions from higher-mass simulations to use as initial conditions.  Although this requires the existence of these higher-mass simulations, a plethora have been run by various groups for various studies; therefore, is it a `simple' matter to mine these simulations for additional data.  As existing proofs-of-concept, several studies of various scales have already investigated where star-forming gas originated \citepeg{LewisBatePrice2015,LewisBate2017,Pelkonen+2021,ArroyocyhavezVazquezsemadeni2022,CollinsLeJimenezvela2023}.  \citet{WursterRowan2023a} likewise determined where star-forming gas originated within the star cluster simulations of \citet{WursterBatePrice2019} and then extracted that gas to construct star-forming clumps.  From an analysis of those initial star-forming clumps, we concluded that stars were born from a wide variety of environments (clumps), where there was a distribution of parameters (e.g., density, Mach number, magnetic field strength and geometry, etc., ...) both within and amongst the clumps.  We concluded that there was not a single universal star forming clump.  Furthermore, we developed those initial clumps such that they could be evolved in future simulations.

In this paper, we evolve 29 star-forming clumps that we extracted from the cluster simulations of \citet[][herein \citetalias{WursterBatePrice2019}]{WursterBatePrice2019} and analysed in \citet[][herein \citetalias{WursterRowan2023a}]{WursterRowan2023a}.  The goal of this study is to determine how closely the evolution of the extracted clumps matches the evolution of the same gas within the cluster simulations themselves.  If the original systems are reproduced, either exactly or statistically, then we submit that the initial clumps as extracted in \citetalias{WursterRowan2023a} provide excellent initial conditions for studies of isolated star formation, since the initial environment is more realistic and we have greatly reduced the number of initial (and free) parameters.  A natural follow-up to this current study will be to re-simulate these clumps at the high-resolution of typical isolated star formation simulations.  This will yield the best of both worlds by permitting star formation to be numerically investigated in high-resolution simulations where the initial conditions are realistic rather than pristine, idealised, and contrived.  

The rest of the paper is organised as follows.  In \secref{sec:methods}, we describe our methods and how they contrast to those in \citetalias{WursterBatePrice2019}.  In \secref{sec:ic} we summarise our initial conditions as generated in \citetalias{WursterRowan2023a}.  We present our results in \secref{sec:results} and provide further discussion in \secref{sec:disc}.  We conclude in \secref{sec:conc}.

\section{Methods}
\label{sec:methods}

We solve the equations of self-gravitating, radiation non-ideal magnetohydrodynamics (MHD) using the three-dimensional smoothed particle hydrodynamics (SPH) code \textsc{sphNG}.  This code originated from \citet{Benz1990}, but is under continual development; it has since been substantially modified to include a consistent treatment of variable smoothing lengths \citep{PriceMonaghan2007}, individual time-stepping and sink particles \citep{BateBonnellPrice1995}, magnetic fields \citerev{Price2012}, and non-ideal MHD \citep{WursterPriceAyliffe2014,WursterPriceBate2016,Wurster2016,Wurster2021} using the single-fluid approximation.  For stability of the magnetic field, we use the \citet{BorveOmangTrulsen2001} source-term approach, and maintain a divergence-free magnetic field using constrained hyperbolic/parabolic divergence cleaning \citep{TriccoPrice2012,TriccoPriceBate2016}.

\textsc{sphNG} has continued to be updated since 2017, when we began the simulations published in \citetalias{WursterBatePrice2019}.  We intentionally chose to use the current\footnote{Current as of the beginning of this project in 2022.} version of \textsc{sphNG} for this study since it includes the most up-to-date algorithms.  We acknowledge that this means that any differences when comparing the current results to those in \citetalias{WursterBatePrice2019} will be a combined result of the extracted cores being isolated from the remainder of the cloud and the different algorithms.  We considered using the same version of \textsc{sphNG}, however, science should always move forwards, and we want our results to be as current as possible, even with the reduced ability to compare to the source data.  The three main differences are as follows.

In this study, we model radiation as flux-limited diffusion \citep{WhitehouseBateMonaghan2005,WhitehouseBate2006}, whereas \citetalias{WursterBatePrice2019} used the combined radiative transfer and diffuse interstellar medium model of \citet{BateKeto2015}.  The latter method uses flux-limited diffusion for the optically thick dense regions and a thermodynamics model for the optically thin, low density material.  While this worked well in \citetalias{WursterBatePrice2019} which includes a component of the interstellar medium (ISM), when we tested it on small clumps in the current study, the thermodynamics model substantially and artificially heated the boundaries of the clump, causing it to dissipate.  Since the clumps (and cores) in the original model would be shielded from the ISM, we are justified in using only the flux-limited diffusion model.  We note that the \citet{BateKeto2015} model may work for our larger clumps, but we intentionally chose to use the same radiation algorithm for the entire suite.  

In this study, we use the artificial resistivity algorithm outlined in \citet{Price+2018phantom} rather than the algorithm presented in \citet{TriccoPrice2013}.  The current algorithm is less resistive \citep{Wurster+2017}, thus the resulting discs are less influenced by numerical processes.  In ideal MHD simulations, we thus expect to form smaller discs in the current study due to less artificial resistivity. 

In this study, we use v2.1 of \textsc{Nicil} \citep{Wurster2021} rather than v1.2.1 \citep{Wurster2016} to calculate the non-ideal coefficients.  The new version uses a reduced chemical network, whereas the previous version used two proxy elements.  In general, resistivity is slightly higher in v2.1 than v1.2.1, suggesting slightly larger discs should form in the non-ideal clumps in the current study.

\section{Initial conditions}
\label{sec:ic}

The method to generate our initial conditions is described in detail in \citetalias{WursterRowan2023a}.  Briefly, at the end each simulation in \citetalias{WursterBatePrice2019} (i.e., at $t_\text{final} = 276$~kyr), we identify the existing sink particles.  We then identify all the particles associated with each sink, where we define an associated particle to be either bound to the sink or within a critical radius of $r_\text{min} = 50$~au. Moving backwards in time, we repeat this process and continually associate new particles with the sink as necessary.  Once a dump file is reached where the star does not exist (i.e., the dump file before it formed), then we create a `pseudo-sink'; we add all the particles in the pseudo-sink to the list, and add close and bound particles one final time.  We then track these particles back in time until the maximum density decreases to \rhoxeq{-16}; at this time, we augment these particles with all the particles within 4$h$ of any associated particle, where $h$ is the smoothing length of a particle.  This is defined as our clump, and the time this occurs is defined as the extraction time, $t_\text{extract}$.  The centre of mass is shifted to the origin and the velocity is shifted to remove any bulk velocity.  

We then embed the clump into a warm background medium whose density is 30 times lower than the average density at the boundary of the clump.  The thermal energy of the background is set so that, on average,  the background and clump are in pressure equilibrium.  Finally, the magnetic field of the background is uniform and is equal to the average magnetic field of the boundary; this action is performed component-wise.  The box itself is periodic, and is approximately twice the size of the clump in each direction.   

As in \citetalias{WursterBatePrice2019}, we include sink particles with accretion radius $r_\text{acc} = 0.5$~au.  When a gas particle reaches \rhoeq{-5} near the end of the second collapse phase, it is replaced with a sink particle.  All particles within 0.25~au of the centre of the sink are automatically accreted, and all particles within 0.5~au are tested to determine if they meet the criteria to be accreted. Since we resolve the opacity limit, each sink particle represents one star or brown dwarf and the properties of the sink reflect the properties of the star. 

For this study, we maintain the numerical resolution of \citetalias{WursterBatePrice2019}, which is $10^{-5}$~\Msun{} for each equal-mass SPH particle.  

Each clump is evolved for a length of time equal to $t_\text{final} - t_\text{extract}$, therefore the total runtime of each simulation is different.  Due to limited computational resources, for this study, we chose to investigate the clumps extracted from models \nfive{} and \ifive{} in \citetalias{WursterBatePrice2019}, which were initialised with a normalised mass-to-flux ratio of $\mu_0 = 5$; this value matches the mass-to-flux ratio used in many of our previous studies of isolated star formation \citepeg{WursterPriceBate2016,WursterBatePrice2018sd,WursterBatePrice2018hd,Wurster+2022}.  Therefore, we evolve the 17 clumps extracted from \nfive{} and the 12 clumps extracted from \ifive{}, which, as with the parent simulation, are evolved employing non-ideal MHD and ideal MHD, respectively.  We explicitly note that there was a stellar merger in \ifive{}, therefore, the clump construction process of 11 clumps began at \tfinal{}, while the construction process of the final star began at $t < $ \tfinal{}.  

For reference, \figref{fig:nfive:end} shows gas cloud in model \nfive{} from \citetalias{WursterBatePrice2019} at two different scales at \tfinal{}.
\begin{figure*} 
\centering
\includegraphics[width=0.5\textwidth]{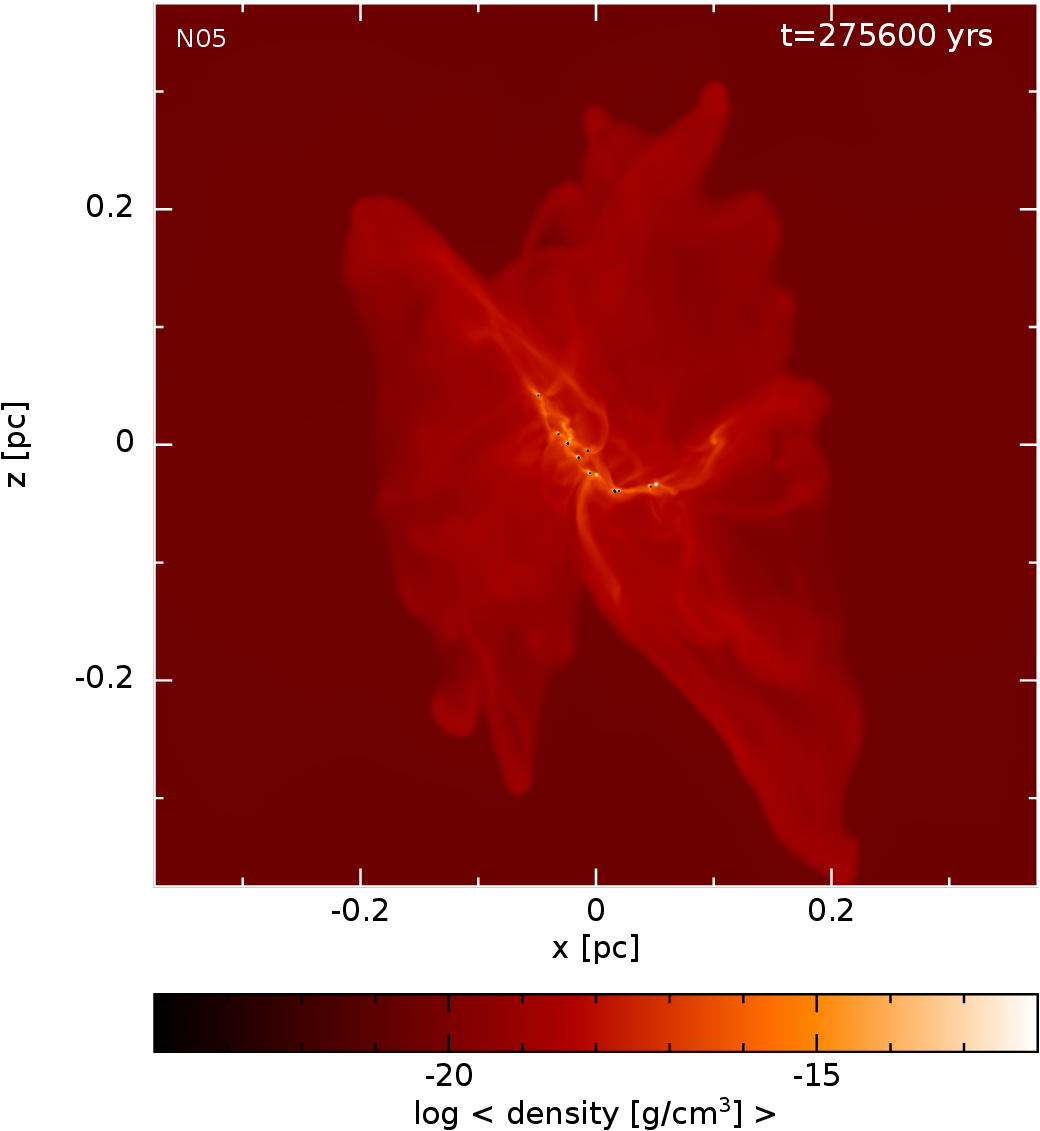}
\includegraphics[width=0.5\textwidth]{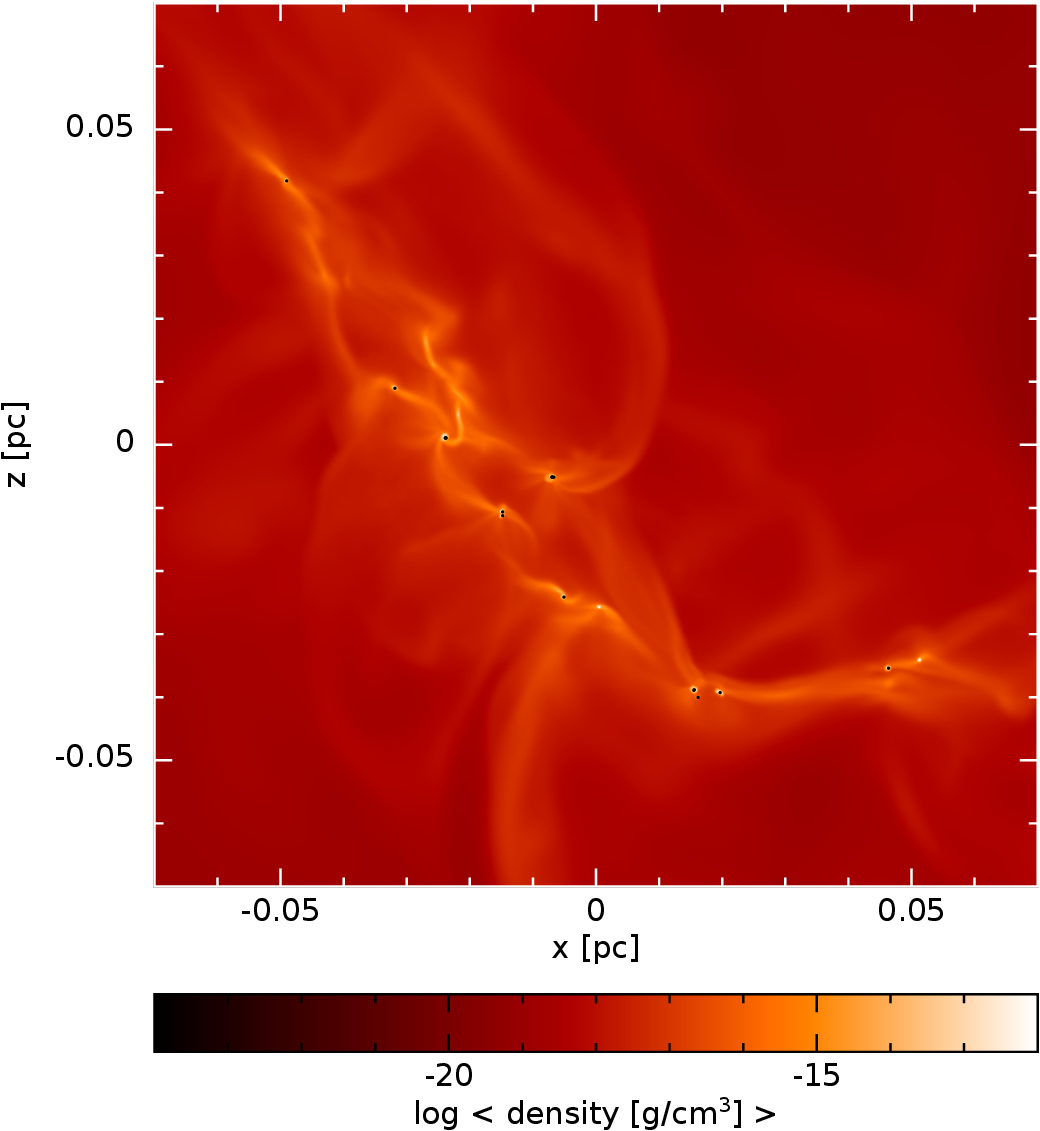}
\caption{Average gas density of cloud \nfive{} from \citetalias{WursterBatePrice2019} at two different scales at the final time of $t_\text{final} = 276$~kyr.  The black dots represent sink particles, and have a radius of 350 and 100 times the accretion radius of the sink for the top and bottom panel, respectively.  The cloud  extends to $r \sim 0.2$~pc (top), however, the dense structures and star formation are confined to a region of $r \sim 0.07$~pc (bottom). No stellar region exists in isolation, thus any region that has contracted has had its gas replenished.}
\label{fig:nfive:end}
\end{figure*} 

\subsection{Definitions}
\label{sec:ext:def}
For simplicity and clarity, we re-state the important definitions here that we will use throughout the paper; these are the same as in \citetalias{WursterRowan2023a}.
\begin{itemize}
\item \textit{Parent simulation} or \textit{parent cloud}:  The simulations presented in \citetalias{WursterBatePrice2019}.  Each cloud initially contained 50~\Msun{} of gas.
\item \textit{Progenitor sink particle} or \textit{progenitor star}: The sink particle in the parent simulation that was used to construct the clump.
\item \textit{Progenitor system}: The hierarchical system to which the progenitor star belongs at \tfinal{}. This is to help guide the analysis and is not used in the construction of the clumps.
\item \textit{Associated clump}: The complete list of SPH particles that will ultimately accrete onto the sink, come within the critical distance of the sink, or become bound to the sink; associated clumps were created in \citetalias{WursterRowan2023a}, but are only used in this section of this paper.
\item \textit{Extracted clump} or \textit{clump}: The final list of SPH particles that includes the associated particles plus all their neighbours.  This definition does not include the background particles.  This is the clump to which we are referring throughout the rest of this paper.  This is referred to as an \textit{augmented clump} in \citetalias{WursterRowan2023a}.
\item \textit{Extraction time}: The time at which the maximum density in the associated clump decreases to \rhoxle{-16}, defining the initial location of the associated clump.
\item \textit{N05}: The name of the parent simulation in \citetalias{WursterBatePrice2019} that was initialised with $\mu_0 = 5$ and evolved using non-ideal MHD.
\item \textit{I05}: The name of the parent simulation in \citetalias{WursterBatePrice2019} that was initialised with $\mu_0 = 5$ and evolved using ideal MHD. 
\end{itemize}

\subsection{Simulations names}
The clumps are named as \textit{X}c\textit{Y}, where \textit{X} is the name of the parent simulation and  \textit{Y} is a number based upon the order in when the clump was extracted, from earliest to latest.

The progenitor systems at \tfinal{} are identified as \textit{X} [\textit{Z}].   At $t_\text{final}$, \nfive{} has four isolated stars, one binary system, one triple system, and two systems of four stars which are identified as $Z  = $ (s1, s2, s3, s4, b1, t1, q1, q2), respectively.  At the same time, \ifive{} has five isolated stars and three binary systems which are identified as $Z  = $ (s1, s2, s3, s4, s5, b1, b2, and b3+), respectively; the latter includes a plus sign since a star previously merged with a star in that system.  Note that all these systems are dynamically evolving; they form before \tfinal{} but are not guaranteed to survive beyond \tfinal{}.

\section{Results}
\label{sec:results}

\subsection{General evolution and large-scale structure}
\label{sec:results:evol}
\figref{fig:evol} shows the evolution of six clumps extracted from \nfive{}.
\begin{figure*} 
\centering
\includegraphics[width=\textwidth]{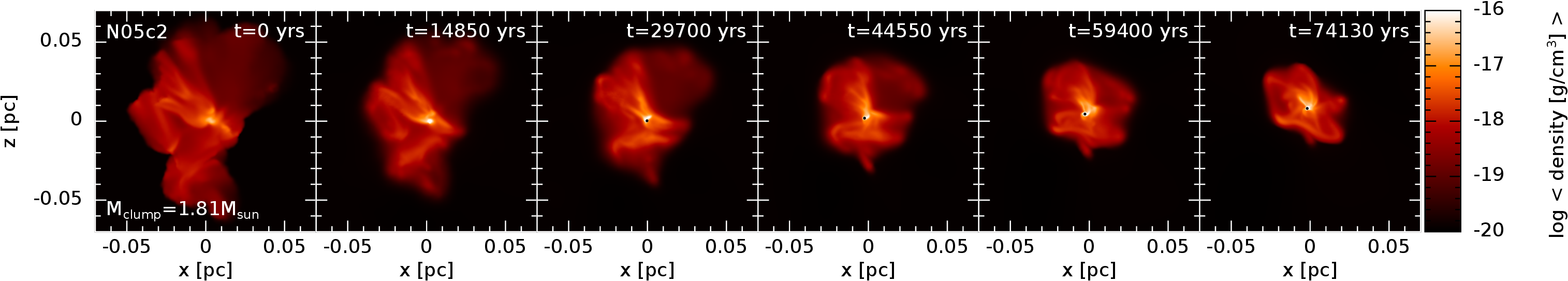}
\includegraphics[width=\textwidth]{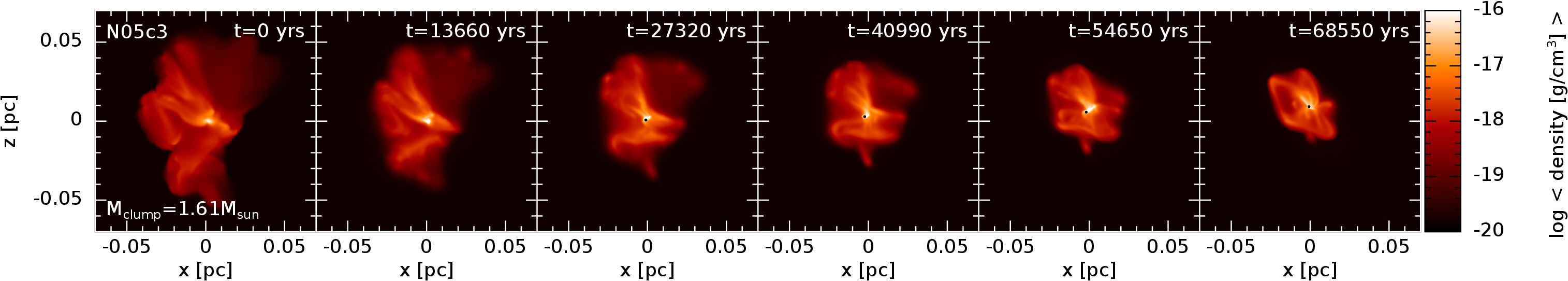}
\includegraphics[width=\textwidth]{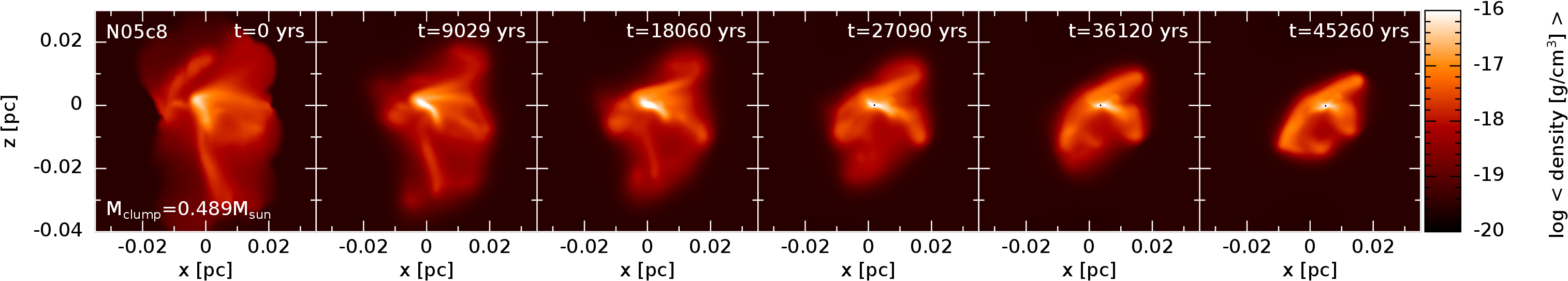}
\includegraphics[width=\textwidth]{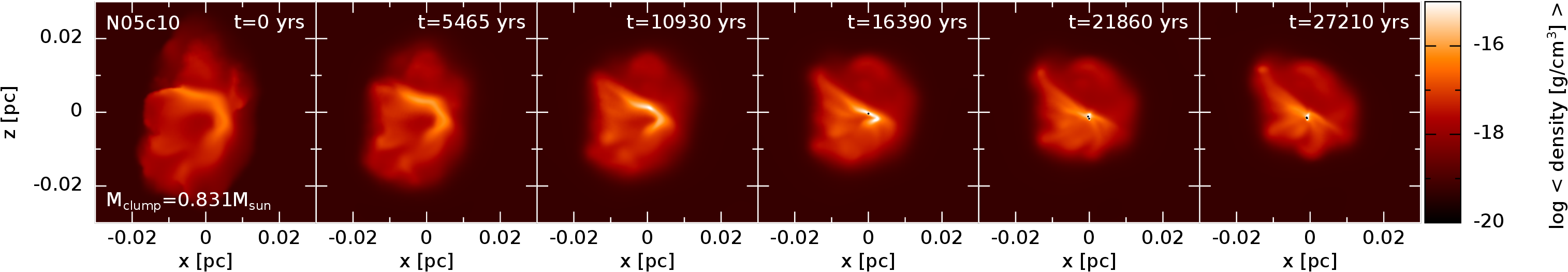}
\includegraphics[width=\textwidth]{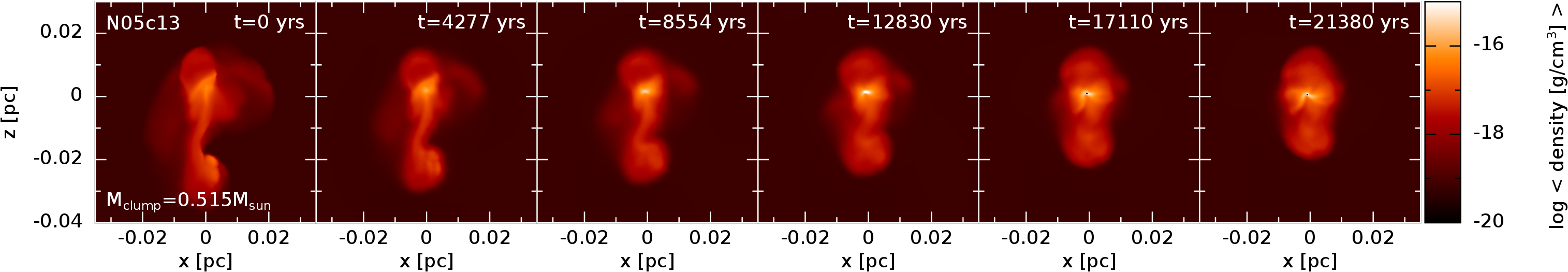}
\includegraphics[width=\textwidth]{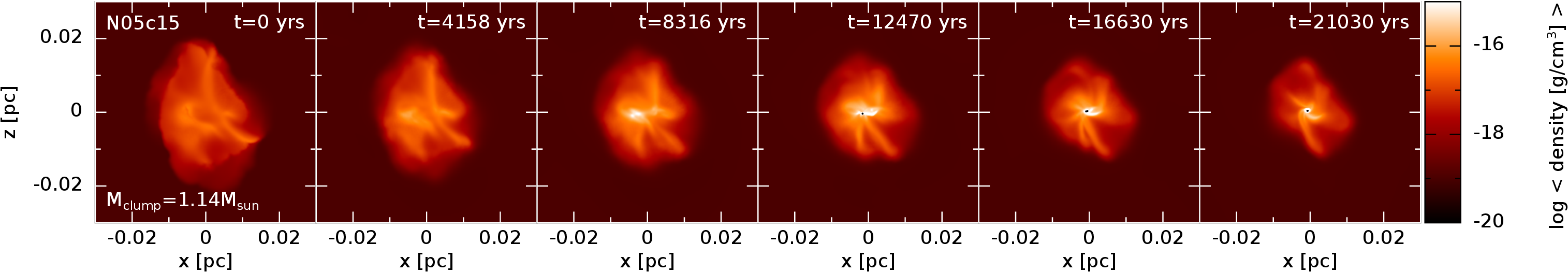}
\caption{The evolution of six clumps extracted from \nfive{}, with the time of extraction increasing from top to bottom.  The times in each panel are zeroed to the beginning of the evolution of the respective clump.  The black dots represent the sink particles and have radii 350 and 100 times that of the accretion radius for the top two and bottom four rows, respectively.  Note that the rows use different spatial and density scales to better highlight the features.  Star formation proceeds rapidly (within \sm10-20~Myr) as the entire clump gravitationally contracts.}
\label{fig:evol}
\end{figure*} 
Since most stars in \citetalias{WursterBatePrice2019} are in hierarchal systems at $t_\text{final}$, it is unsurprising that some initial clumps appear similar since gas can be associated with several progenitor stars (first column; see also \figref{fig:ics} and fig. A1 in \citetalias{WursterRowan2023a}).  As a demonstration, we have intentionally selected \cnfive{2} and \cnfive{3} where the progenitors are in a tight binary system (and part of the widely separated quadruple system \snfive{q1}).  The top two rows show the global evolution of \cnfive{2} and \cnfive{3}, where the two evolutions appear qualitatively similar, particularly the structures near the dense core.  The most notable difference is that \cnfive{3} is smaller at all times, which can be accounted for by its lower initial mass (1.61 compared to 1.81~\Msun{}).  Moreover, 1.60~\Msun{} of gas is common to both clumps, therefore, \cnfive{3} is effectively a subset of \cnfive{2}.

When comparing \cnfive{10} and \cnfive{11} where the progenitors were also in a binary system (\snfive{b1}, which is less tight than above), the two clumps demonstrate nearly identical large-scale evolution (the latter is not shown).  In this case, the clumps were extracted at the same time with a very large overlap in SPH particles (0.75~\Msun{} of gas is common, where the clump masses are 0.83 and 0.78~\Msun{} for \cnfive{10} and \cnfive{11}, respectively).  

Therefore, since nearly all the hierarchal systems in \citetalias{WursterBatePrice2019} were formed by capture, the difference in evolution of clumps with progenitors from the same system depends on the initial separation of the progenitor stars, and how much different gas they interacted with.   The tables in \appref{app:overlap} show the percentage of common gas amongst clumps whose progenitors are in a bound system at \tfinal{}.  It is also more likely that younger clumps spawned from the same progenitor system will have similar evolutions compared to older clumps since there is less time for their histories to diverge.

Despite the initial turbulence, the evolution here proceeds similarly to that of simulations of isolated star formation.  Here, the maximum gas density continually increases throughout the first hydrostatic core phase until the first star is formed within \sm10-20~kyr.  During this time, the clump is globally gravitationally collapsing and its volume is decreasing.   Although this is in agreement with simulations of isolated star formation, it disagrees with the results of \citetalias{WursterBatePrice2019} where star forming clumps or cores never detach from their surroundings (c.f. \figref{fig:nfive:end}).  This disagreement suggests that our results will be minimally impacted by the new numerical algorithms discussed in \secref{sec:methods}.

For a direct comparison between our extracted clumps and the region around the progenitor star for the six clumps, see \figref{fig:nfive:compare:big}.
\begin{figure*} 
\centering
\includegraphics[width=\textwidth]{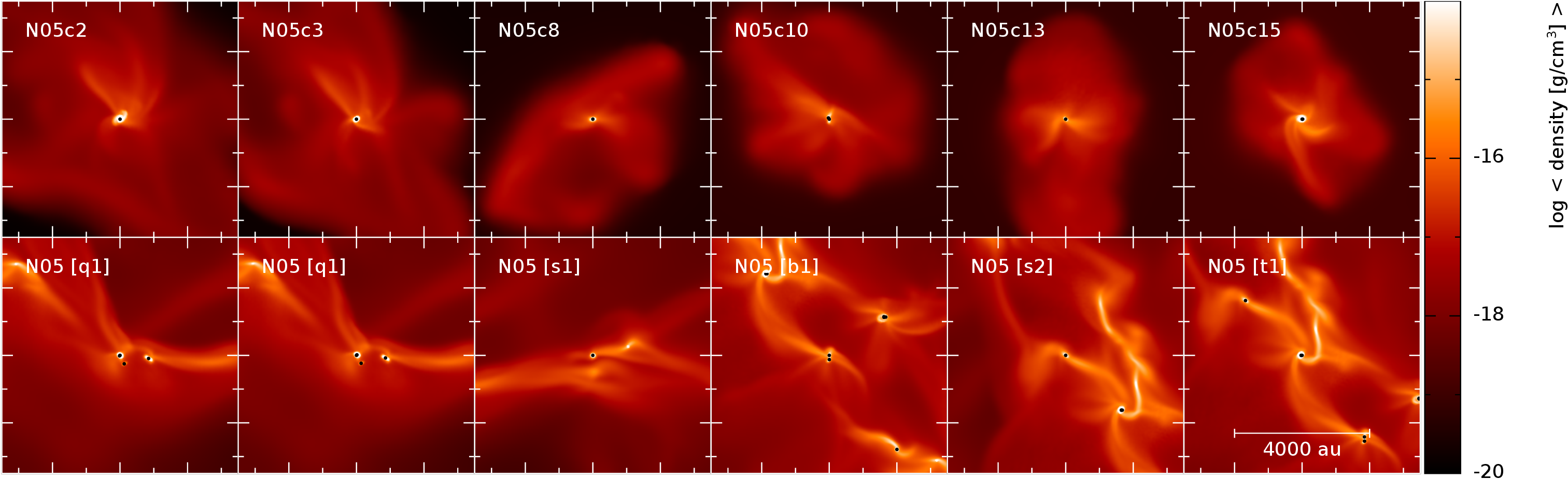}
\caption{The average gas density in the $xz$-plane of six clumps at the final time (top; centred on the first star to form) compared to the region from the parent simulation centred on the progenitor star (bottom).  The black dots represent the sink particles, and have a radius 100 times that of the accretion radius.  The large-scale filamentary structure present in the parent simulations is not reproduced by the evolving the extracted clumps.}
\label{fig:nfive:compare:big}
\end{figure*} 
At this final time, the parent clouds (bottom) are filamentary, with most of the stars located within the filaments or at intersections of filaments; the progenitor to \cnfive{15} is being fed by at least two different filaments.  However, the surrounding gas in the extracted clumps (top) does not contain the same well-defined filamentary structure.  There are a few potential striations feeding \cnfive{2} and \cnfive{3}, and slightly better defined striations feeding \cnfive{15}; the latter are over-densities that are rotating about the star and only slowly feeding the star rather than well-defined filaments.  In all cases, the striations in the extracted clumps extend from the centre to the edge of the clump, suggesting that they are shaped by larger-scale processes than contained within our clumps.  Therefore, our extraction method as developed in \citetalias{WursterRowan2023a} \textit{does not} reproduce the large-scale filamentary structures, or even the localised filamentary structures around the final star(s).  

\subsubsection{Stellar evolution as a proxy for clump evolution}
To quantitatively evaluate the evolution of the clumps compared to their progenitors, we will use stellar (sink) evolution as a useful proxy.   \figsref{fig:sink:evolN05}{fig:sink:evolI05} show the mass evolution of the individual stars in the clumps and the progenitor systems for \nfive{} and \ifive{}, respectively, where the horizontal axis spans the evolved lifetime of the clump.  
\begin{figure*} 
\centering
\includegraphics[width=\textwidth]{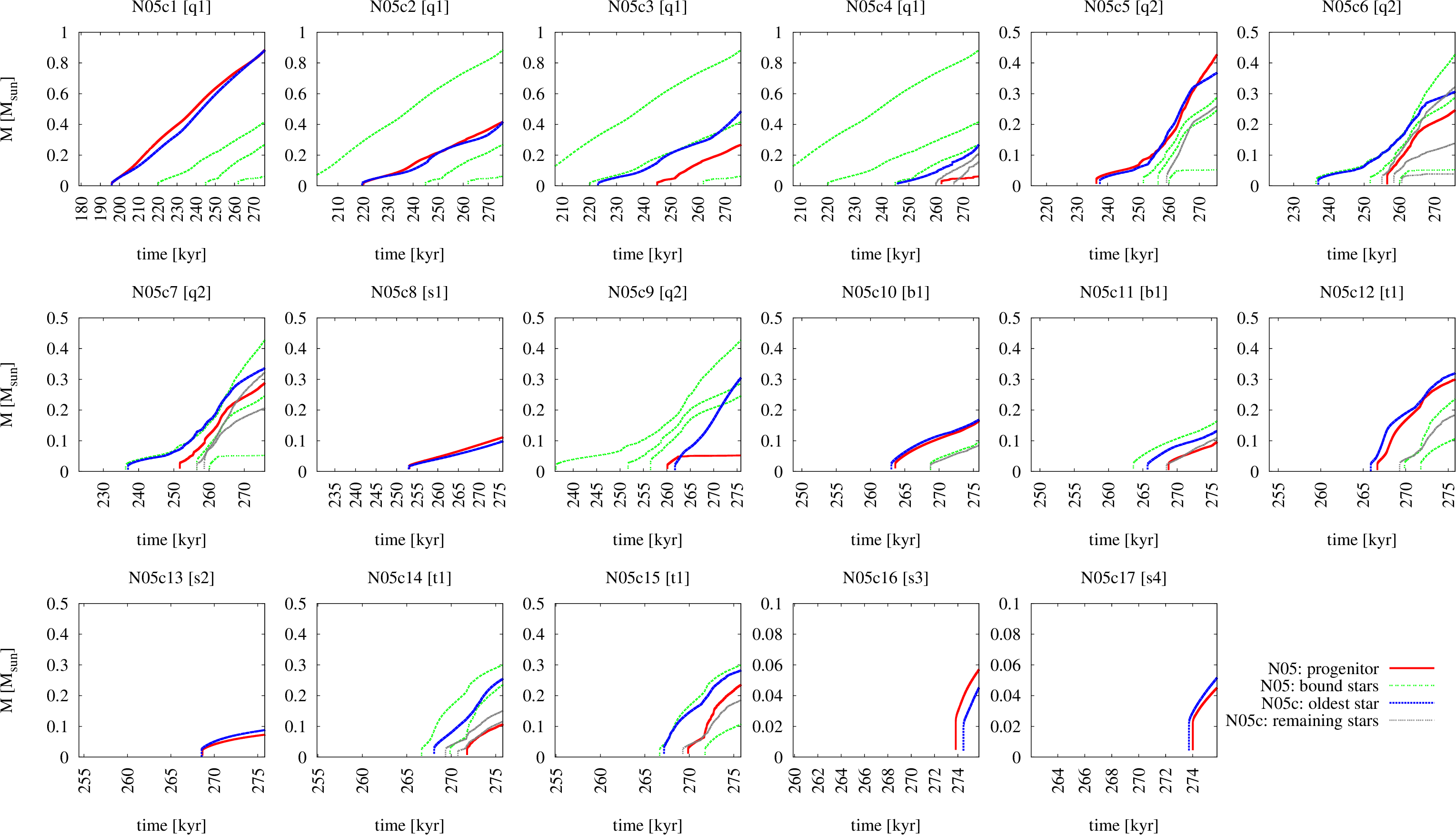}
\caption{Mass evolution of the stars in all 17 clumps extracted from \nfive{} and from \nfive{} itself.  In each panel, we compare the evolution of the progenitor star (red) and the remaining stars in the progenitor system (green) to the first star to form in the clump (blue) along with the remainder of its stars (grey). The time axis is zeroed to the beginning of the evolution of the parent cluster \nfive{}, thus the lines representing the stars in the extracted clumps are shifted by the extraction time of the clump.  The time axes spans from the extraction time of the clump to \tfinal{}, thus is different in each panel; the vertical axis varies to best highlight the stellar mass evolution.  The primary star in the clumps typically follows the mass evolution of one star in the progenitor system.  The masses and mass evolution of individual stars in the clumps appears to better agree with the progenitor systems than when comparing the discs and disc systems.}
\label{fig:sink:evolN05}
\end{figure*} 
\begin{figure*} 
\centering
\includegraphics[width=\textwidth]{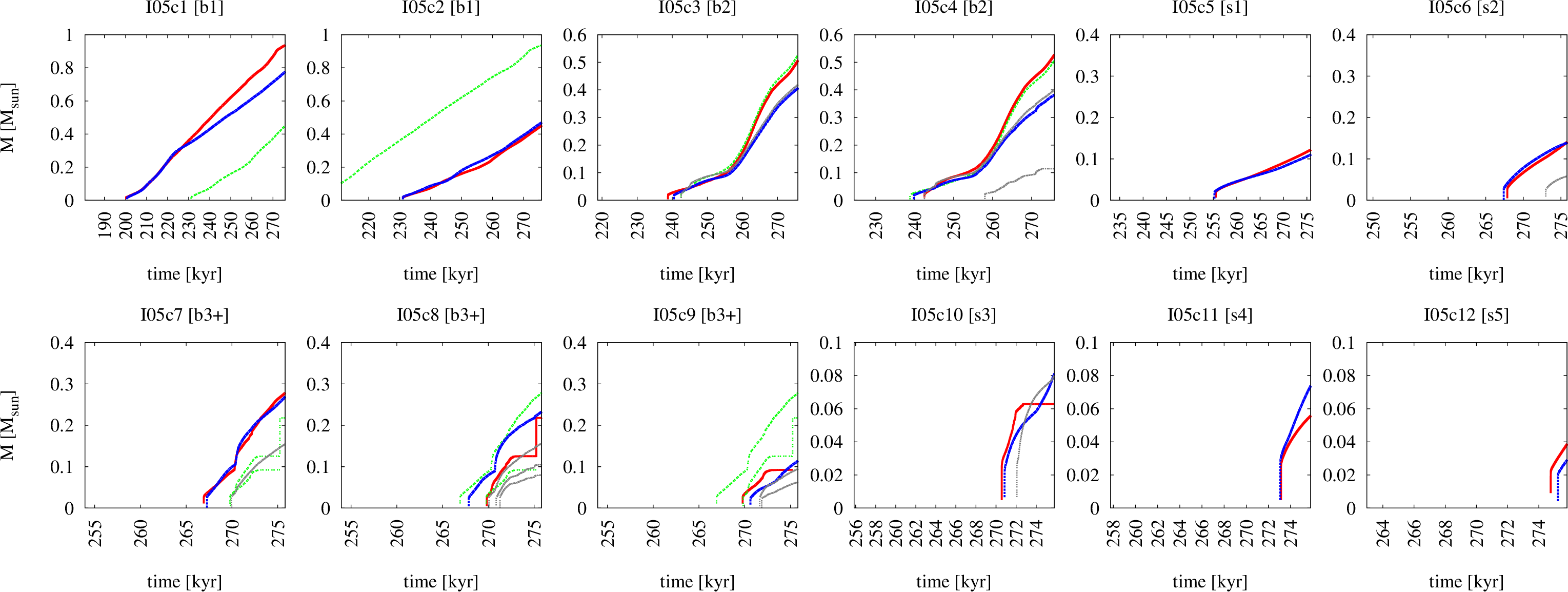}
\caption{Mass evolution of the stars in all 12 clumps extracted from \ifive{} and from \ifive{} itself, as in \figref{fig:sink:evolN05}.  The legend is the same as in \figref{fig:sink:evolN05} and given only in that figure for clarity.}
\label{fig:sink:evolI05}
\end{figure*} 
If the stellar evolution is reproduced exactly, then in each panel, the red and blue lines will match; if the stellar system evolution is reproduced exactly, then additionally the green and grey lines will also match.  We clearly see that we do not identically reproduce any system.

In most cases, all the stars in a progenitor system form after the extraction time of a given clump.  However, there are exceptions, such as the progenitors to \cnfive{9},  \cnfive{1}, and  \cnfive{2}.  The first case represents a star that rapidly proceeds from nearby diffuse gas to birth and only interacts with the low-density gas of the circumtriple disc, while the latter two cases represent stars that for much of their life are unbound thus do not interact with the same gas. 

In most cases in the clumps extracted from \nfive{}, the star in the extracted clump forms before its progenitor and has more mass than its progenitor star; there is no clear trend for the stars that form in the clumps extracted from \ifive{}.  In most cases in all models, the first star to form in the extracted clump reasonably well follows the mass evolution of one star in the progenitor system.  In some cases (e.g., \cnfive{1,2,8,10}), this is the progenitor star itself, suggesting that our extraction algorithm can reproduce that star.  However, in many cases (e.g. \cnfive{3,6,15}), the first star follows the evolution of a different star in the progenitor system.  In many of these latter cases, the oldest star in the clump follows the evolution of an older star in the progenitor system, suggesting that its core is at least partially included in the clump and is already somewhat evolved at extraction time of the clump. 

In most cases, the evolutionary tracks of the clump and cloud stars are similar, suggesting a similar accretion mechanism in all simulations, at least on the scales near the star itself.  One notable star, however, is the progenitor to \cnfive{9}.  This star is ejected from its system in the parent simulation, and its accretion is halted.  In the clump simulation, however, only a single star forms and it does not need to compete with additional stars for material.  These two evolutions demonstrate the effect of dynamical interaction and the competition for material.  It is likely that if the progenitor never interacted with the three other stars in \nfive{}, it would also have grown to a large mass.

In summary, the first star to form in the clump has a similar evolution to one of the stars in the progenitor system.  Therefore, our extraction algorithm permits us to reasonably reproduce the accretion history of one star in the progenitor system, even if it is not the progenitor star.  However, we are less able to reproduce the accretion history of an entire progenitor system. 

\subsection{Final states}
\label{sec:final}
In \figsref{fig:evol}{fig:nfive:compare:big}, we have shown how the large-scale gas evolution differs between the clumps and the parent clouds, in part due to the non-existence and existence of filaments, respectively.  As previously stated, this is a result of the extraction method in \citetalias{WursterRowan2023a} extracting clumps based upon the gas that was associated with a progenitor star and disc and not the surrounding filaments.  Despite the lack of filaments, our algorithm design meant we were able to reasonably reproduce the mass evolution of some stars (\figsref{fig:sink:evolN05}{fig:sink:evolI05}).  However, how well is the progenitor system (including discs) reproduced?  \figsref{fig:nfive:compare}{fig:ifive:compare} compare the stars and discs of all the clumps extracted from \nfive{} and \ifive{}, respectively, to their progenitors at the end of the simulation. 
\begin{figure*} 
\centering
\includegraphics[width=\textwidth]{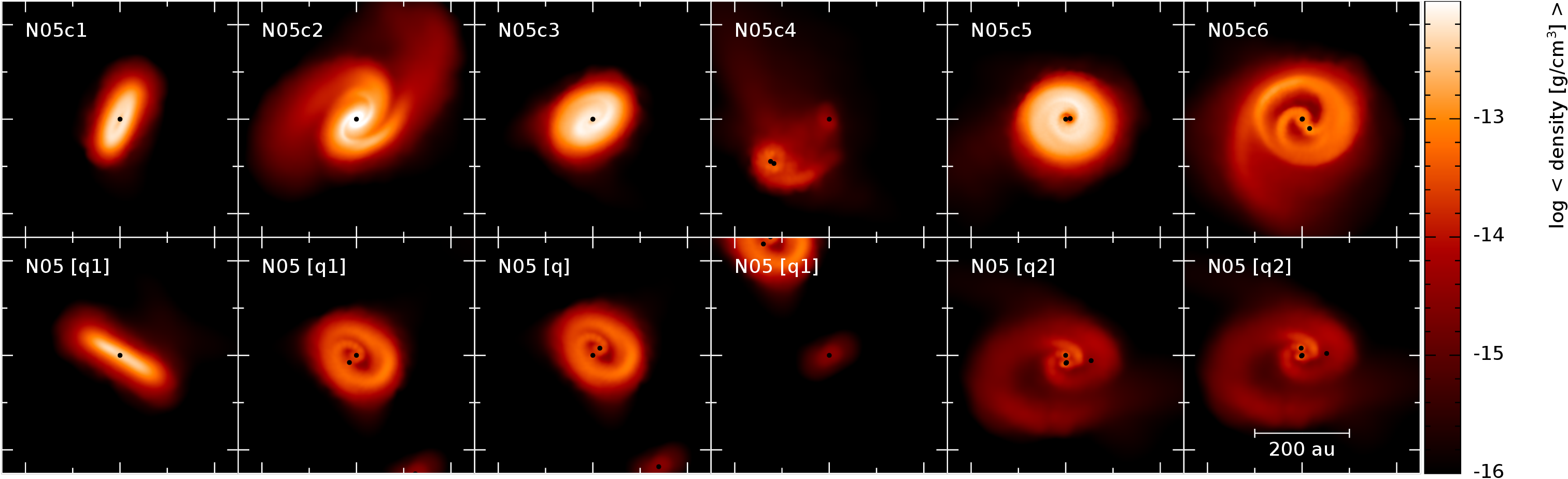}
\includegraphics[width=\textwidth]{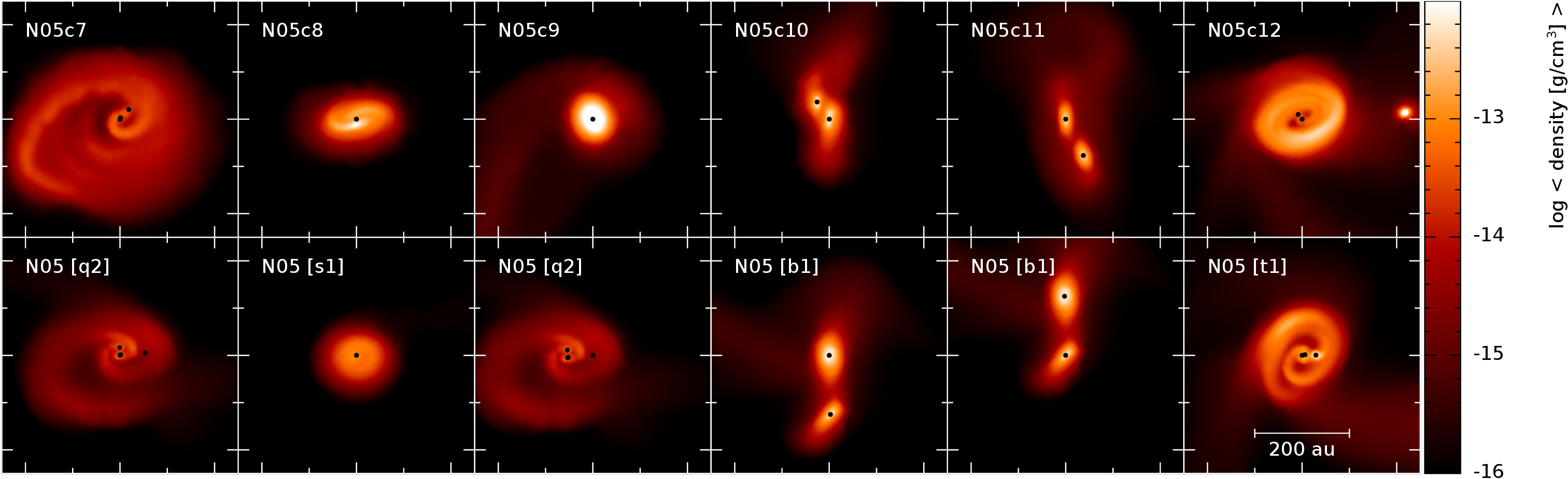}
\includegraphics[width=\textwidth]{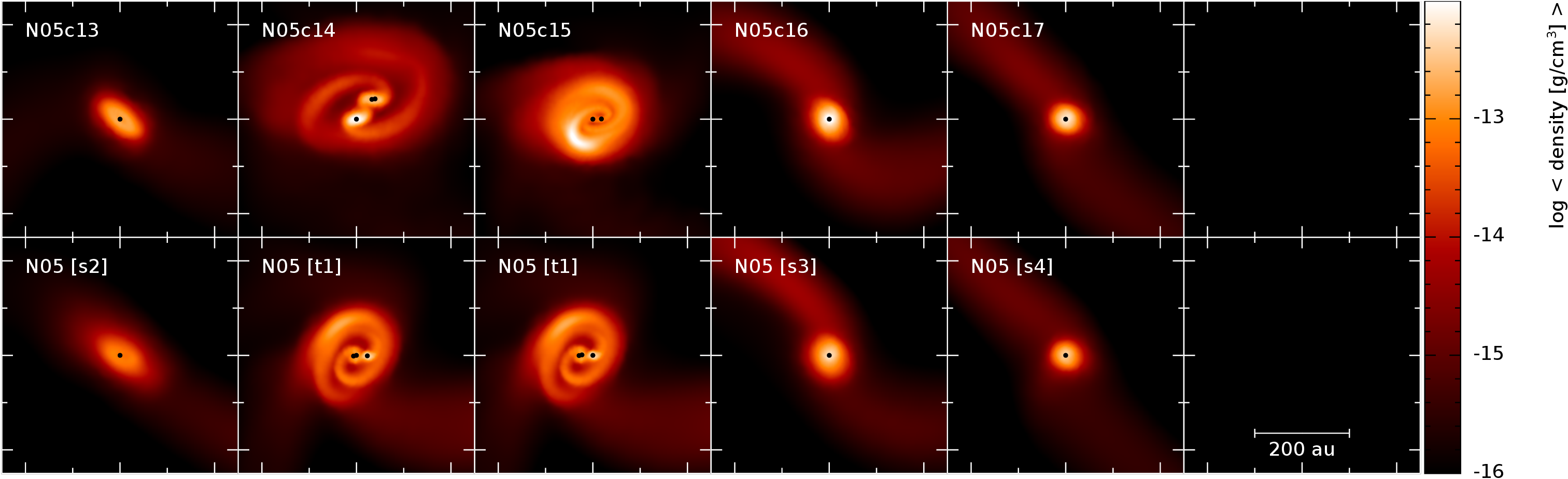}
\caption{The average gas density in the $xz$-plane of all 17 clumps extracted from \nfive{} at the final time (top row of each panel; centred on the first star to form) compared to the region from the parent simulation centred on the progenitor star (bottom row of each panel).   The black dots represent the sink particles and have a radius 10 times that of the accretion radius.  The systems that form from the clumps reproduce their progenitor systems with varying degrees of accuracy, with better accuracy for the progenitor low-order systems.}
\label{fig:nfive:compare}
\end{figure*}
\begin{figure*} 
\centering
\includegraphics[width=\textwidth]{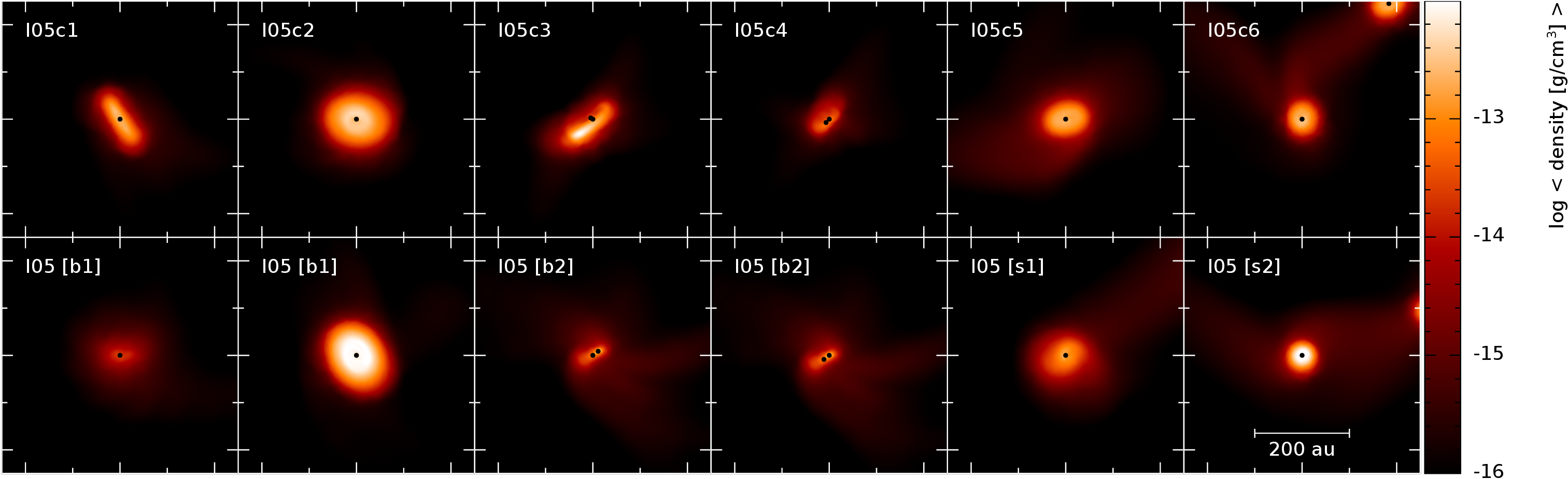}
\includegraphics[width=\textwidth]{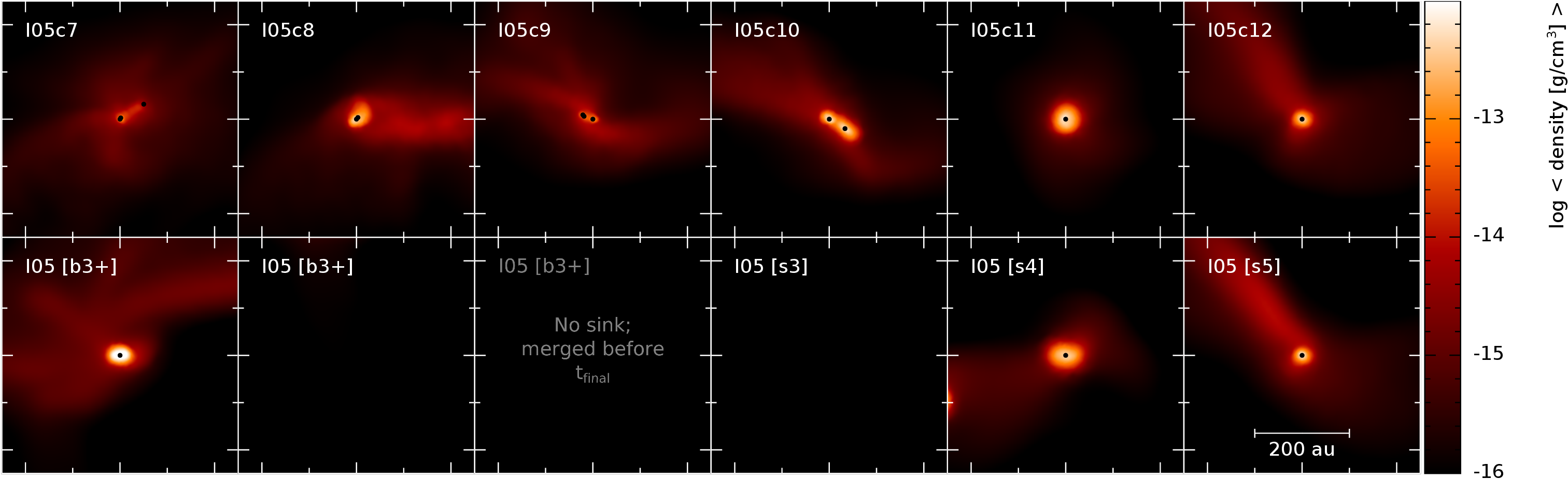}
\caption{The average gas density in the $xz$-plane of all 12 clumps extracted from \ifive{} at the final time (top row of each panel; centred on the first star to form) compared to the region from the parent simulation centred on the progenitor star (bottom row of each panel).   The black dots represent the sink particles and have a radius 10 times that of the accretion radius.  The progenitor to \cifive{9} merged with the progenitor to \cifive{8} prior to the final time; we have evolve this clump to $t_\text{final}$ but have not included a comparison image.  The progenitors of \cifive{8} and \cifive{10} are present in the middle of their respective panels, but cannot be seen since the gas density is too low.  The systems that form from the clumps reproduce their progenitor systems with varying degrees of accuracy, with better accuracy for the progenitors with well-defined discs.}
\label{fig:ifive:compare}
\end{figure*} 
Although some systems are in good qualitative agreement with their progenitor systems, our clumps clearly do not form a one-to-one reproduction of their progenitor.

For a quantitative comparison between the stars and discs from the extracted clumps to the progenitor stars and discs, \figref{fig:disc:Properties} compares the stellar system mass, disc mass, disc-to-star mass ratio, the disc radius, and the magnetic field strength of the discs.
\begin{figure*} 
\centering
\includegraphics[width=\textwidth]{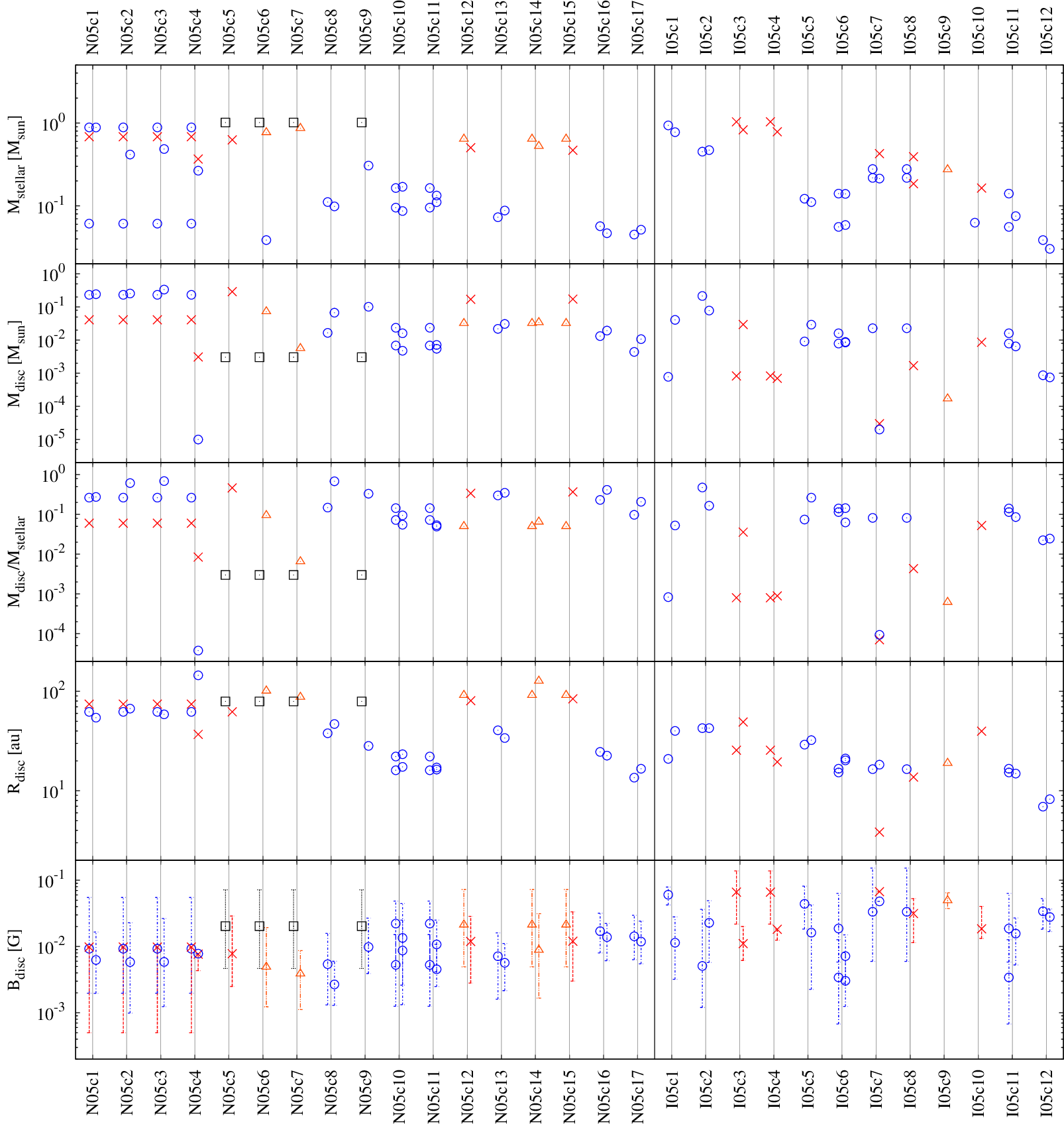}
\caption{From top to bottom: mass of the stellar system,  mass of the disc, ratio of the disc mass to the stellar mass, disc radius enclosing 63.2 per cent of the mass, and magnetic field strength of the disc where the bars span 95 per cent of the field strengths in the disc.  The circles represent circumstellar discs, the x’s represent circumbinary discs, the triangles represent circumsystem discs about three stars, and squares represent circumsystem discs about four stars.  Although multiple discs exist per system, often hierarchically, we have selected the discs that best represent the system (e.g., see \figrref{fig:nfive:compare:big}{fig:ifive:compare}).  The symbols to the left of the vertical lines represent stars and discs in the parent clouds while the symbols to the right represent stars and discs in the extracted clumps.  \cifive{6} and \cifive{11} have progenitors that are only 426~au apart but not bound, however, we have included both progenitors for the comparisons of \cifive{6} and \cifive{11}.  The progenitor to \cifive{9} merged prior to the end of the simulation, hence the lack of points on the left-hand side of the line.  There is a varying degree of agreement amongst the models, but this suggests that discs properties may be reasonably reproduced even if their morphological features are not (e.g., see \figsref{fig:nfive:compare}{fig:ifive:compare}).  }
\label{fig:disc:Properties}
\end{figure*} 
In this figure, we plot the properties of the highest-order disc(s) rather than the highest order system.  For example, \snfive{q1} is a quadruple system, but clearly there is no circumsystem disc about all four stars; thus, we instead plot the properties of the two circumstellar discs and one circumbinary disc.  On the left-hand side of each thin grey line, we plot the properties of the stars and discs associated with the progenitor star (system) from the parent simulation, and on the right-hand side, we plot the properties of the stars and discs from the extracted clumps.  

From \figrref{fig:nfive:compare:big}{fig:ifive:compare}, some clumps reasonably reproduce their progenitor systems, although never exactly.  There is typically better agreement when the progenitor is in a lower-order system (i.e., isolated or a binary) extracted at a later time (i.e., is a younger clump).  Thus for young low-order systems, it appears that the clump extraction method of \citetalias{WursterRowan2023a} is reasonable to reproduce the system.   However, given the preceding results and figures, in general, it is clear that we cannot extract gas and have it robustly and identically evolve as when it is embedded in the parent cloud.  Given the chaos and non-linearity of star formation, and the presence of large-scale processes, this is not surprising. 

In the remainder of this section, we discuss the stellar and disc properties and distributions at the final time.  We must be cautious about small-number statistics since we are only evolving 17 (12) non-ideal (ideal) MHD clumps, not all of which form discs.  We also note that these are dynamically evolving systems, so the time of analysis may also play a role in our conclusions.  In the following histograms, we do not normalised the distributions, and all histograms have bin widths of 0.5~dex.  We plot the parent distribution twice, once where each object in the parent simulation is counted only once (parent-U) and once where we repeatedly count stars and discs each time they appear in a progenitor clump (parent-RC); i.e., every point in \figref{fig:disc:Properties} is included once in the parent-RC histogram.   The mean and standard deviation of the histograms for the non-ideal and ideal simulations are given in Tables~\ref{tab:dist:nimhd} and \ref{tab:dist:imhd}, respectively.  We also compare each pair of histograms within each magnetic prescription using the  two-sample Kolmogorov-Smirnov test (K-S test) and present the $p$-value in these tables; values larger than 0.05 suggest that the two distributions are compatible with being drawn from the same (but unknown) distribution. 

\begin{table*}
\tiny
\begin{center}
\begin{tabular}{l c c c c c c c c c c}
\hline
                                &  N: pU & N: pRC & N: clump & $\mu\pm\sigma$: pU & $\mu\pm\sigma$: pRC& $\mu\pm\sigma$: clump & K-S: pU \& pRC & K-S: pU \& clump & K-S: pRC \& clump \\
\hline 
Stellar mass              & 17 & 49 & 31 & 0.16\Msun{}$\pm$0.37dex  & 0.20\Msun{}$\pm$0.37dex  & 0.19\Msun{}$\pm$0.30dex & 0.95 & 0.47 & 0.67 \\
Stellar system mass & 11 & 27 & 21 & 0.18\Msun{}$\pm$0.50dex  & 0.29\Msun{}$\pm$0.51dex  & 0.23\Msun{}$\pm$0.43dex & 0.90 & 0.67 & 0.025 \\
Disc mass                & 10 & 23 & 20 & 0.018\Msun{}$\pm$0.51dex  & 0.023\Msun{}$\pm$0.61dex  & 0.025\Msun{}$\pm$1.0dex & 0.96 & 0.31 & 0.33 \\
Disc radius              & 10 & 23 & 20 & 37.9\Rsun{}$\pm$0.28dex  & 49.4\Rsun{}$\pm$0.27dex  & 44.8\Rsun{}$\pm$0.30dex & 0.70 & 0.92 & 0.44 \\
Disc magnetic field  &10 &23 & 20 & 0.025G$\pm$0.47dex  & 0.037G$\pm$0.47dex & 0.010G$\pm$0.32dex & 0.76 & 0.003 & 2.3$\times10^{-6}$ \\
\hline
\end{tabular}
\caption{The properties of the distributions for the non-ideal models.  In the header, pU (pRC) represents parent-U (parent-RC) for brevity.  The average and standard deviation are calculated using the logarithm of the values for consistency with the histograms.   The final three columns are the $p$-values from the two-sample K-S test, where values above 0.05 suggest that the two distributions are compatible with being drawn from the same distribution.  }
\label{tab:dist:nimhd} 
\end{center}
\end{table*}
\begin{table*}
\tiny
\begin{center}
\begin{tabular}{l c c c c c c c c c c}
\hline
                                &  N: pU & N: pRC & N: clump & $\mu\pm\sigma$: pU & $\mu\pm\sigma$: pRC& $\mu\pm\sigma$: clump & K-S: pU \& pRC & K-S: pU \& clump & K-S: pRC \& clump \\
\hline 
Stellar mass              & 11 & 22 & 23 & 0.20\Msun{}$\pm$0.43dex  & 0.22\Msun{}$\pm$0.39dex  & 0.15\Msun{}$\pm$0.35dex & 1 & 0.16 & 0.046  \\
Stellar system mass & 10 & 15 & 15 & 0.19\Msun{}$\pm$0.47dex  & 0.20\Msun{}$\pm$0.46dex  & 0.22\Msun{}$\pm$0.42dex & 1 & 0.94 & 0.89\\
Disc mass                & 8 & 12 & 14 & 0.0065\Msun{}$\pm$0.81dex  & 0.0067\Msun{}$\pm$0.73dex  & 0.0027\Msun{}$\pm$1.1dex & 1 & 0.44 & 0.31 \\
Disc radius              & 8 & 12 & 14 & 19.3\Rsun{}$\pm$0.22dex  & 18.9\Rsun{}$\pm$0.19dex  & 20.3\Rsun{}$\pm$0.29dex & 1 & 0.91 & 0.53\\
Disc magnetic field & 8 & 12 & 14 & 0.099G$\pm$0.39dex  & 0.092G$\pm$0.41dex & 0.022G$\pm$0.25dex & 1 & 0.001 & 0.0004\\
\hline
\end{tabular}
\caption{The properties of the distributions for the ideal models. }
\label{tab:dist:imhd} 
\end{center}
\end{table*}

\figref{fig:histo:sink} shows the distribution of stellar masses for both non-ideal and ideal MHD simulations.  
\begin{figure} 
\centering
\includegraphics[width=\columnwidth]{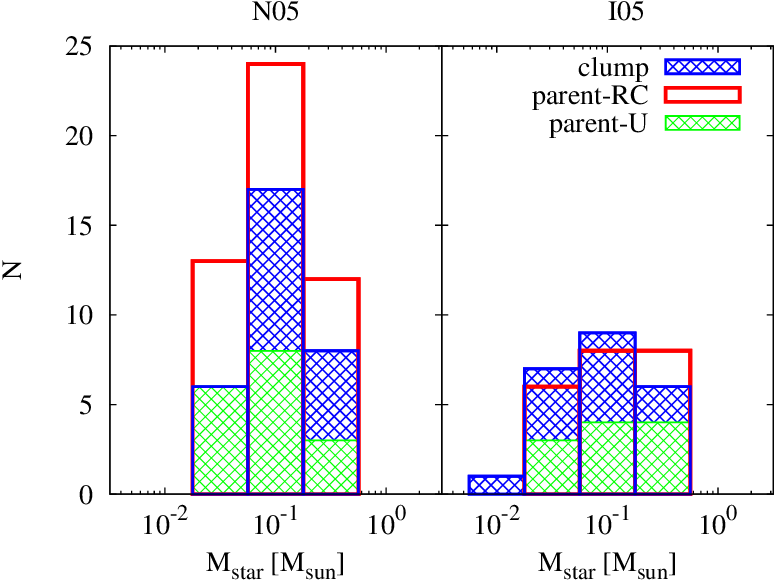}
\caption{Histogram showing the distribution of stellar masses in the parent (red and green) and clump (blue) distributions.  For the red parent model (parent-RC), we repeatedly count stars such that they are included once for every progenitor system in which they appear, while the green distribution (parent-U) includes each star only once.  All distributions span a similar range and are qualitatively similar, despite each distribution having a difference number of members. }
\label{fig:histo:sink}
\end{figure} 
The parent-RC and clump distributions for the non-ideal MHD models each have a different number of members, with 49 and 31 stars, respectively.  Given how we have double counted the parent population, this shows that we have unintentionally suppressed star formation in our extracted clumps, therefore, we have removed some of the material or large-scale processes that promote star formation in the parent cloud (e.g., inertial inflows).  However, this conclusion is unique to the non-ideal MHD models since both the parent-RC and clump distributions of the ideal MHD models have nearly the same number of stars, with 22 and 23 stars, respectively.  Given these two contradictory results, we cannot offer a conclusion on whether or not star formation is suppressed in the clumps compared to the parent clouds. 

Independent of the number of members in the non-ideal MHD distributions, they all span a similar range and are qualitatively similar, with the $p$-value from the K-S test suggesting that they are compatible with being drawn from the same distribution.  The ideal MHD clump distribution is slightly narrower than the parent distributions, and the K-S test suggests that the clump and parent-RC distributions are not  compatible with being drawn from the same distribution, although the clump and parent-U distributions are.  This weaker and non-agreement for parent-U and parent-RC, respectively, is likely due to the small number statistics and the clumps distribution containing one lower mass star while parent-U (parent-RC) contains one (two) higher mass star(s); this reinforces the caution of our quantitive results due to our small numbers.   This shows that our extraction process reasonably and statistically reproduces the stellar distribution for non-ideal MHD, and, in the clumps, we do not form stars that are either much more or much less massive than in the parent cloud; the average stellar mass is lower in the ideal MHD clumps than their parents, however, it is statistically reproduced when comparing with parent-U.  Therefore, our extraction process yields clumps that can be evolved to statistically reproduce the parent-U distribution of stellar masses, despite stars from each clump not exactly forming stars with the same masses as in the progenitor system, as discussed with the evolutionary tracks shown in \figsref{fig:sink:evolN05}{fig:sink:evolI05}.  

When we instead look at the total stellar mass in a system (see \figref{fig:histo:stellar}), we see that the non-ideal MHD distributions are qualitatively similar, except that the parent-RC distribution also has several more massive systems.  However, this most massive bin contains the system \snfive{q2} in quadruplicate, thus skewing the conclusion, which is readily seen when comparing to the parent-U distribution.  The results of our K-S test show that the clump and parent-U distributions are compatible with being drawn from the same distribution whereas, due to the inclusion of the most massive system four times in parent-RC, the clump and parent-RC distributions are not compatible with being drawn from the same distribution.  The top row of \figref{fig:disc:Properties} reinforces that the difference between the clump and parent distributions is from the older, higher order systems (i.e., \snfive{q1,q2}); however, the slightly younger triple system, \snfive{t1}, has reasonable agreement with the stellar masses, even if the clump and parent systems do not contain the same number of stars.  Thus, age of the system is likely the primary contribution to any disagreement between the distributions, followed by multiplicity.

For the ideal MHD models, we again see reasonable agreement between the distributions and the K-S test again concludes that the clump distribution is compatible with being drawn from the same distributions as the parent-U and parent-RC distributions.  Additionally, \figref{fig:disc:Properties} shows reasonable agreement for most of the systems.  Unlike the non-ideal MHD models, there is greater disagreement from the `middle-aged' systems \sifive{7-11} rather than the older systems; multiplicity is not high enough for a comment. 

Therefore, from the distributions and a direct qualitative comparison, it is reasonable to conclude that we can reproduce the stellar system mass from our extraction algorithm in a statistical sense; deviations in direct comparisons between the clumps and parent typically increases with age and multiplicity of the system.  Again, we offer caution due to small numbers and greater impact of outliers.  
\begin{figure} 
\centering
\includegraphics[width=\columnwidth]{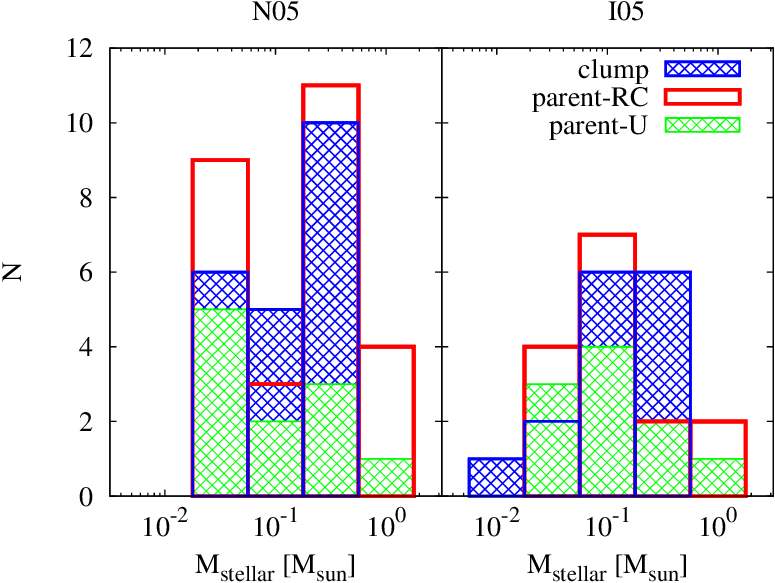}
\caption{Histogram showing the distribution of stellar system masses for the parent  (with repeated-counting in red; with unique counting in green) and clump (blue) distributions.  The parent clouds can form slightly more massive systems, however, this is not a robust conclusion given the small number of systems in each distribution and how system masses were double-counted in the parent distribution.}
\label{fig:histo:stellar}
\end{figure}

The disc mass distributions are shown in \figref{fig:histo:disc}.  There is more deviation between the parent and clump distributions, both in shape and spread, than compared to the stellar and stellar system masses; for these distributions, the standard deviation can be $\lesssim$ dex.  Despite this, all four $p$-values are greater than 0.05.   There is better agreement when comparing the distributions of their radii (\figref{fig:histo:Rdisc}); all four $p$-values for the disc radii are greater than those for the disc masses.
\begin{figure} 
\centering
\includegraphics[width=\columnwidth]{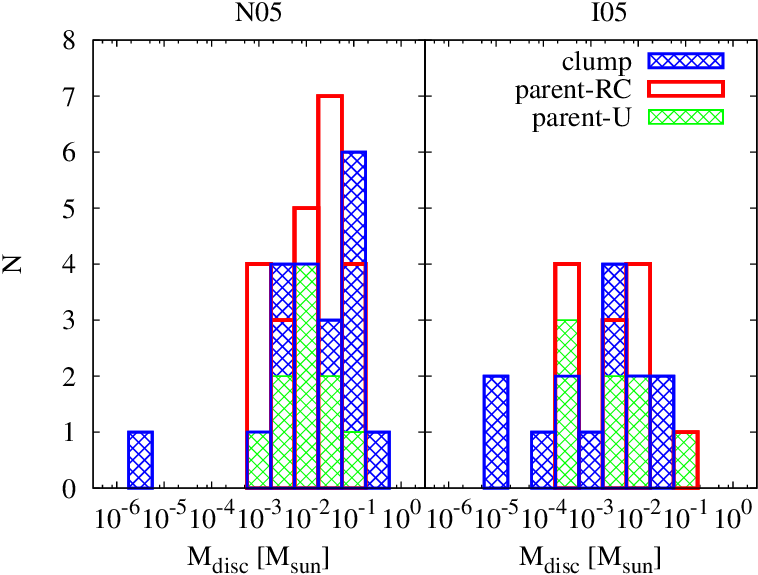}
\caption{Histogram showing the distribution of disc masses for the parent (with repeated-counting in red; with unique counting in green) and clump (blue) distributions.  The clumps yield a wider distribution of disc masses, and contain larger discs since clumps have fewer large-scale processes that can truncate a disc.}
\label{fig:histo:disc}  
\end{figure} 
\begin{figure} 
\centering
\includegraphics[width=\columnwidth]{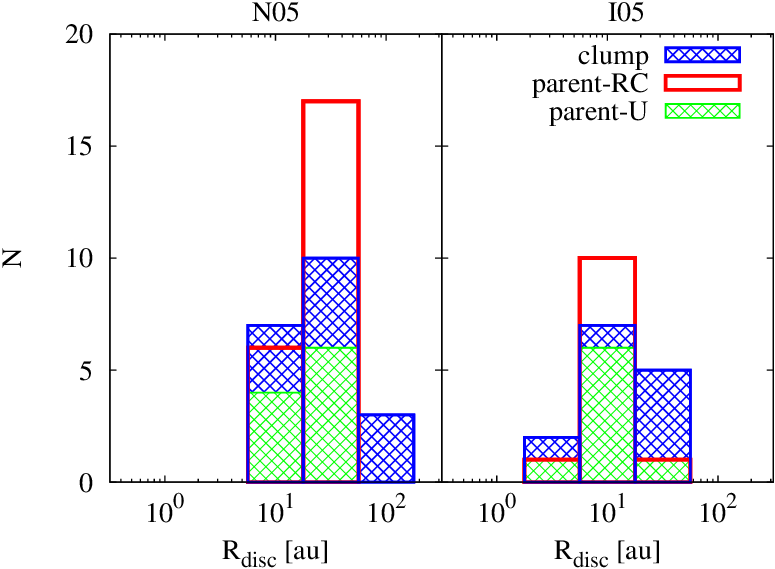}
\caption{Histogram showing the distribution of disc radii for the parent  (with repeated-counting in red; with unique counting in green) and clump (blue) distributions.  Both parent and clump distributions are similar, although the clumps contain larger discs since clumps have fewer large-scale processes that can truncate a disc. }
\label{fig:histo:Rdisc}
\end{figure} 
This is reinforced in \figsref{fig:nfive:compare}{fig:ifive:compare} which renders the final discs in the evolved clumps to their progenitors.  These K-S values are reflected in the comparison of the disc masses and radii shown in the second and fourth rows of \figref{fig:disc:Properties}, respectively.  When comparing the resulting disc mass and radius to their progenitors, there is generally good agreement between the radii for both the ideal and non-ideal MHD models, whereas there can be a factor of \sm10 between the disc masses (although this can be a few dex in the extreme cases).

Although our extraction process includes all the gas that becomes bound to the parent star (which also includes all the gas in the progenitor disc), it does not include the gas that may interact but never becomes bound.  Therefore, both the clump and the parent cloud contain different SPH particles that will interact differently and hence have different effects on the disc morphology.  In addition to different external interactions, our distributions are dependent on the particular snapshot in time since dynamical interactions are continually stripping and feeding the discs.  Therefore, the distribution of discs properties is a less reliable comparison than stellar properties.

Despite the qualitative differences, for both the disc mass and radius distributions, the K-S test suggests that the clump distribution is compatible with being drawn from the same distributions as the parent-RC and parent-U distributions.

Finally, \figref{fig:histo:Bdisc} shows the distributions of average magnetic field strengths in the discs at the final time.  For both ideal and non-ideal MHD, there is a systematic shift that the discs in the clumps have a lower magnetic field strength; the average field strength is \sm3 times lower for the non-ideal MHD models and \sm4 times lower for ideal MHD.  All four $p$-values are less than 0.05, indicating that the clump distributions are not compatible with being drawn from the same distributions as either parent distributions.   This shift is reasonable for the non-ideal MHD models since the clumps were evolved with a newer, slightly more resistive version of \textsc{Nicil}.  However, all clumps were evolved with the new artificial resistivity algorithm that is less resistive suggesting that the discs in the clumps (at least for ideal MHD) should be slightly stronger.  Therefore, it is possible that the clump or cloud size has a direct effect on the evolution of the large-scale magnetic field and thus ultimately how that field affects the field in the discs; exploring this further is out of the scope of the current work.  Given our previous comments about chaotic evolution of the discs, and the different numerical algorithms governing artificial and physical resistivity, we cannot reach any robust conclusions regarding the magnetic field in extracted versus progenitor discs other than to say that the clump distribution is not representative of the parent distributions. 
\begin{figure} 
\centering
\includegraphics[width=\columnwidth]{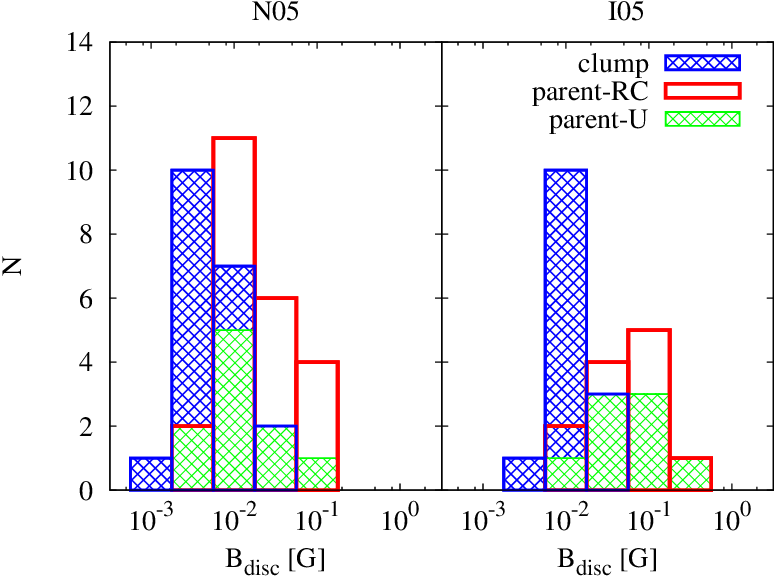}
\caption{Histogram showing the distribution of disc magnetic field strengths for the parent (with repeated-counting in red; with unique counting in green) and clump (blue) distributions.  The distributions are systematically lower in the disc clumps than the parent clumps and are not drawn from the same distributions as per the results of our K-S test.  We cannot determine the cause given that disc evolution is chaotic and that the clumps and parent clouds were evolved with different physical and artificial resistivity algorithms.}
\label{fig:histo:Bdisc}
\end{figure} 

In most non-ideal MHD cases, the average field strength is only slightly lower in the clump than its progenitor, with the exception of \snfive{q2}.  However, in all non-ideal MHD cases, the ranges of magnetic field strengths in the discs have a large overlap between the clump and progenitor discs; see the bottom row of \figref{fig:disc:Properties}, where the vertical bars span 95 per cent of the field strengths in the discs.  The magnetic field strength in most discs spans \sm dex, although there is an even larger spread in \snfive{q1}, which promotes the overlap.  Despite the overlap, our previous comments about the effect of the improved physical and artificial resistivity comments remain true.  There is more variation amongst the average and range of magnetic field strengths in the ideal MHD discs, both within the progenitor and clump disc distributions and between the two distributions.  This reinforces that non-ideal MHD processes help to regulate the evolution of the magnetic field.  

Given the influence that the magnetic field strength has on the disc properties in simulations of isolated star formation \citepeg{WursterPriceBate2016,LewisBate2018,WursterBate2019}, it is surprising that there is statistical agreement amongst disc radii and masses while the magnetic fields are much weaker in the discs; however, the importance of the magnetic field on disc properties is reduced in cluster simulations (e.g.,  \citealp{Seifried+2013}; \citetalias{WursterBatePrice2019}), suggesting that the evolution of the clumps is more analogous to that of star formation in clusters rather than in isolation.  For a better understanding of the effect of the magnetic field, future studies should test the evolution of the clumps using the older versions of the physical and artificial resistivity algorithms; for further discussion on the effect of physical and artificial resistivity, see \citet{Wurster2016,Wurster2021} and \citet{Wurster+2017,WursterBatePrice2018ff}, respectively.  Therefore, this suggests that our decision to use the most up-to-date algorithms as discussed in \secref{sec:methods} may have been incorrect.

\section{Discussion}
\label{sec:disc}

As discussed in \secref{sec:results} and shown in \figref{fig:evol}, well-defined filaments do not form in our clumps.  The gas is gravitationally collapsing on to the dense core, although in some cases, this flow is redirected by a striation rather than retaining a somewhat radial collapse.  In other cases, the gas retains or gains angular momentum and spirals around the core before accreting onto it.  These gas flows are different than in the parent clouds.  

Several studies of isolated low-mass star formation have included decaying turbulence \citepeg{Seifried+2012,Seifried+2013,Joos+2013,LewisBate2018,WursterLewis2020d,WursterLewis2020sc,Hennebelle2021}.  All of these results include structured remnants, where the filamentary flows are better defined for studies with higher rms Mach numbers and higher initial masses.  All of our initial cores are supersonic with $1.5 \lesssim \mathcal{M} \lesssim 3$, however, they are also low-mass with $M_0 \lesssim 2$~\Msun{}.  Despite the supersonic Mach number and that the parent cloud formed well-defined filaments, our extracted clumps behave like isolated star forming clumps.  Therefore, whether modelling isolated star formation in idealised initial cores or in clumps extracted from larger-scale simulations, processes on even larger scales are required to generate and sustain filaments.  Future studies are required to investigate the gas flow in both the parent simulation and the extracted clumps to determine the impact and importance of the filaments, and to determine why stars with similar mass distributions can be formed with and without the presence of filaments.

From our analysis of the extracted clumps, it is clear that, even when accounting for all the gas that is ever associated with a progenitor star, the original system cannot be identically reproduced.  This includes the multiplicity, the stellar masses, and the discs.   Therefore, large-scale processes are irrevocably lost during the extraction procedure.  Additionally, even when evolving two similar clumps (e.g., \cnfive{2} and \cnfive{3}, where the former includes nearly all the gas in the latter plus an additional \sm12 per cent), we see different evolutions and final states.  It has been previously and repeatedly shown that different initial realisations of turbulence produce different results \citepeg{Liptai+2017,Geen+2018}, and we can extend this conclusion by stating the even slightly different subsets of gas from a parent simulations will produce different results; however, in all cases, the slightly different realisations or subsets yield results that are statistically similar.  This reinforces that star formation is a non-linear, chaotic process and any sort of change (including removing so-called `unimportant' gas from a simulation or slightly changing an initial turbulent field) will prevent exact reproducibility.  However, star formation may be statistically robust to these small changes.

How gas flows differ between the clump and the parent cloud suggests that different star formation mechanisms may be prevalent on different scales.  On large scales, gas is funnelled towards the stars through the filaments, thus suggesting the inertial inflow model \citep{Padoan+2020}; in agreement, in \citetalias{WursterRowan2023a}, we argued that star formation was likely proceeding via the internal inflow model in the parent cloud itself, and that a clump was never a well-defined, isolated region.  On the system scale, the multiple stars and dynamical interactions suggest that the stars are competing for the local gas, which is evidence for the competitive accretion model \citep{Zinnecker1982,Bonnell+2001ca,Bonnell+2001acc}; this is notable in \cnfive{9} which does not need to compete for gas and grows to 0.3~\Msun{} compared to its progenitor which does need to compete for gas and is ejected after failing to do so.  Therefore, we postulate that star formation (or rather the formation of clumps where stars will ultimately form) proceeds via the inertial inflow model on the large-scale, while stars grow via competitive accretion \textit{within} a clump.

\subsection{Hydrodynamics}
The pure hydrodynamic cloud (named \hyd{}) in  \citetalias{WursterBatePrice2019} formed 19 stars; these progenitor stars yielded several massive clumps \citepalias{WursterRowan2023a}, up to $\gtrsim20$ per cent of the parent cloud.  Therefore, evolving these hydrodynamic clumps is much more computationally expensive than evolving the clumps extracted from \nfive{} and \ifive{}, despite the absence of magnetic fields.  Since the goal of \citetalias{WursterRowan2023a} is to produce low(ish)-mass clumps that will form lower-order systems that can ultimately be run at higher resolution, our goal was not met with most of the clumps in \hyd{}.

Nonetheless, we evolved a few of the hydrodynamic clumps.  We found that star formation proceeded very efficiently, and too many stars were formed to perform a useful comparison as above.  This is a result of the reasonably high stellar density in \hyd{}, which meant that most progenitor stars interacted with several other stars and large quantities of gas, yielding massive clumps.  Given the mass of these clumps, it is unsurprising that they behave more like clusters and produce many stars than clumps like and produce few stars. 

Therefore, a limitation of our extraction algorithm is that it cannot well reproduce stellar systems in clumps that have been extracted from a dense stellar cluster.  Given the weak magnetic fields in models \ntwenty{} and \itwenty{} (non-ideal and ideal MHD models, respectively, initialised with normalised mass-to-flux ratios of 20) from \citetalias{WursterBatePrice2019}, it would be interesting to determine if our method can reproduce those systems given the weak magnetic regulation on star formation in those models.  Answering that question is beyond the scope and resources of this project. 

\subsection{Future work}
The work presented in this paper and in \citetalias{WursterRowan2023a} has many potential spin-off studies, including the following.

\subsubsection{Clump extraction parameters}
As shown and discussed above, our extracted clumps do not form filaments given our extraction algorithm developed in \citetalias{WursterRowan2023a}.  Thus, a next step would be to test the various parameters of that algorithm to determine if we could extract clumps that would from filaments without the influence of the remaining cloud.  Although we expect that this would be possible (e.g., by adjusting our boundness criteria, the radius to which we search for particles, and/or the radius where we unconditionally add particles our clumps), we are not confident that it could be done while continuing to extract reasonably sized ($\lesssim5$~\Msun{}) clumps.  However, this should be explored in future work.

 Additionally, it would be beneficial to explore the effect of extracting the clumps at lower density values (e.g.) \rhoxeq{-17} or at the beginning of the parent simulation.  We can also investigate merging the extracted clumps of stars in a bound system.  In addition to exploring the effect that these changes have on the initial clumps themselves (as in \citetalias{WursterRowan2023a}), it would be important to determine how these clumps evolve and if a set of parameters allows for near identical reproduction of a system, or if all parameters simply yield statistical agreement.  

\subsubsection{Resolution}
\label{sec:disc:res}
We have demonstrated that our extracted clumps statistically reproduce the stellar and discs systems from the parent clusters.  Thus, we have demonstrated that these clumps are excellent initial conditions for simulations of isolated star formation.  Therefore, these clumps can be re-resolved at the higher resolutions typically used in isolated star formation simulations.  Using these clumps as initial conditions will reduce the number of free-parameters required of typically isolated star formation simulations, yielding less contrived results.  Moreover, given the higher resolution, we will be able to investigate features (e.g., outflows) that are not resolved at the resolutions typically used in cluster simulations \citep{Wurster+2022}.  Therefore, running several of these clumps at higher resolution will greatly contribute to our understanding of the star formation process within its chaotic environment.  Although these clumps cannot reproduce filaments using the current extraction algorithm with its default parameters, these clumps are still a preferred initial condition since filaments generally do not form in isolated star formation simulations (e.g., see the papers listed in the Introduction).

An alternative method to generating high-resolution data of stellar systems from less-contrived initial conditions would be to employ live particle splitting \citep{KitsionasWhitworth2002,KitsionasWhitworth2007} in the parent simulations themselves.  This would permit the stellar systems to be resolved at high resolution while still being influenced by the (lower-resolution) larger-scale processes.  However, this could still lead to long run-times.  For example, if we re-resolved 6~\Msun{} of the 50~\Msun{} (e.g., the sum of the masses of the largest clumps extracted from each system in \nfive{}) by splitting each SPH particle into 13 low-mass child particles, then the total number of particles in the parent simulation would increase from \sm$5\times10^6$ to  \sm$12\times10^6$, and the lower-mass particles would have smoothing lengths \sm2.4x smaller than their lower-resolution counterparts.  Between the increased number of particles and smaller smoothing lengths, we could predict an increase in runtime by a factor of \sm5.  Although this is much cheaper than running the entire simulation at high resolution, this would still be prohibitively expensive given the code and the computer architecture the parent simulations were ran on.  An additional consideration to the long run times is the interaction of particles of the same type (e.g., gas) but of different mass (i.e., resolution).   Given the dynamics of the stellar systems, SPH particles will be ejected from the systems, thus an algorithm must be developed to de-resolved these ejected particles to their original resolution.  If not, then the interpenetration of SPH particles of the same type but of different masses will lead to numerical instabilities. 

\subsubsection{Use as initial conditions}
One of the aims of this project is to develop less-contrived initial conditions for simulations of isolated star formation.  Given the qualitative and quantitative results presented here, we argue that our initial conditions are less-contrived than those typically used in studies of isolated star formation and will hence be useful in investigating -- at high resolution as per \secref{sec:disc:res} -- the star formation process and the early evolution of its environment.  We admit that the statistical agreement is not as strong as we would like between the clumps and the parent distributions, however, as discussed in \secref{intro}, most studies in the literature produce a range of results by varying even a single parameter.   Therefore, the results from our clumps are reasonable given the agreement with the parent distributions and the improved algorithm, and is consistent with the literature by producing a variety of results.  

We admit that it is not practical to use lists of particles as initial conditions, particularly for meshed algorithms.  Thus, as a future project, we propose parametrising the initial clumps presented in \citetalias{WursterRowan2023a}.  This future project will aim to produce a suite of star forming cores where each core will have self-consistent parameters (e.g., mass, radius, initial turbulent Mach number, etc...,) and (radial) profiles for (e.g.) density, temperature, velocity, and the magnetic field.  Therefore, it will become a simple selection of parameterised core which can then be readily initialised in any astrophysical code.  This will remove the need to select individual parameters and idealised profiles in order to produce more realistic star formation simulations.

\section{Summary and conclusion}
\label{sec:conc}

In \citetalias{WursterRowan2023a}, we extracted a star forming clump for every star that formed in the low-mass star cluster formation simulation of \citetalias{WursterBatePrice2019}.  In this paper, we evolved and analysed the 17 (12) clumps that were extracted from the star cluster seeded with a normalised mass-to-flux ratio of 5 that employed non-ideal (ideal) MHD.  We maintained the numerical resolution of \citetalias{WursterBatePrice2019}, but we used a newer version of our SPH code to exploit the changes in the six years since that study began.  We evolved the extracted clumps from their extraction time to the final time modelled in \citetalias{WursterBatePrice2019}, thus each clump was evolved for a different length of time.  

Our main conclusions are as follows:
\begin{enumerate}
\item The first star to form in a clump typically follows the mass evolution of a star in the progenitor system, but not necessarily its own progenitor.  The mass accretion rates onto individual stars is similar in both clumps and the parent cloud, suggesting a similar (competitive) accretion mechanism on small scales.  
\item Stellar systems including their discs are not identically reproduced.  Although the clump contains all the gas that will become \textit{bound} to the progenitor star, it does not contain all the gas that will \textit{interact} with the star, thus leading to different evolutions.
\item The distributions of stellar mass and stellar system mass are similar between the clumps and the parent clouds, and we have shown that the clump and parent distributions are drawn from the same underlying distribution.   Therefore, we have statistically reproduced these distributions.
\item The distributions of disc mass and radii are less similar between the clumps and the parent clouds in a qualitative sense, but agree in a statistical sense.  However, discs are dynamically evolving through external interactions, thus this weak (qualitative) agreement is likely a result of a combination of the time we selected to analyse the discs and small number statistics.   
\item The large-scale filamentary structure observed in the parent clouds is not reproduced in the extracted clumps.  Therefore, the production and maintenance of filaments occurs through processes on larger scales than those present in star-forming clumps, even if they are seeded with decaying turbulence.  Despite this, the distributions of the stellar and stellar systems properties are statistically reproduced, suggesting that well-defined filaments may not be responsible for the final properties of a stellar system.  Future investigation is required.
\end{enumerate}

From our analysis, we reiterate the common consensus  that star formation in a chaotic process.  Important physical processes occur on many scales, so if large-scale processes are removed when extracting a clump of gas, then the progenitor system cannot be identically reproduced.  
Therefore, we demonstrated that the gas in the initial clump is continually modified by external gas in the parent system, and that its profile at extraction time is not enough to predict its future evolution.  

We have also shown that systems can be statistically reproduced.  Therefore, we conclude that our goal of generating low-mass clumps that can be used as initial conditions for high-resolution simulations of star formation has been achieved.  Although messier than the traditional, idealised initial conditions, modelling star formation in extracted clumps will provide greater insight into the star formation process and the associated physical processes given the more realistic initial environment.

\section*{Acknowledgements}

We would like to thank the referee for comments that somewhat improved the quality of this manuscript.
We would like to thank Ian A Bonnell for useful discussions.  
JW would also like to thank all his collaborators for many years of helpful and inspiring discussions that have greatly contributed to the quality of his scientific and numerical works; he would like to specifically thank Daniel J Price, Matthew R Bate, Ian A Bonnell, Terrence Tricco, and Rebecca Nealon.  
JW acknowledges support from the University of St Andrews.
CR acknowledges support from the European Research Council (ERC) under the European Union’s Horizon 2020 research and innovation program under grant agreement No 638435 (GalNUC).
This work was performed using the DiRAC Data Intensive service at Leicester, operated by the University of Leicester IT Services, which forms part of the STFC DiRAC HPC Facility (www.dirac.ac.uk). The equipment was funded by BEIS capital funding via STFC capital grants ST/K000373/1 and ST/R002363/1 and STFC DiRAC Operations grant ST/R001014/1. DiRAC is part of the National e-Infrastructure.
Several figures were made using \textsc{splash} \citep{Price2007}.  

\section*{Data availability}

The data underlying this article and \citet{\wbp2019} will be available upon reasonable request.

\bibliography{ResimulationTwo.bib}
\appendix
\section{Initial clumps}
\label{app:overlap}
The top and bottom panels of \figref{fig:ics} respectively show the initial clumps for the 17 and 12 non-ideal and ideal MHD clumps that we evolved.  The clumps are extracted over \sm60~kyr, extracted from a range of regions, and have masses of $0.29 < M_\text{clump}$/\Msun{}$ < 2.11$.   As discussed in \secref{sec:results:evol}, there are similarities between clumps extracted from the same progenitor system, however, these clumps are not identical and can be extracted over an extended period of time (e.g.,  28~kyr for \snfive{q1}).
\begin{figure*} 
\centering
\includegraphics[width=\textwidth]{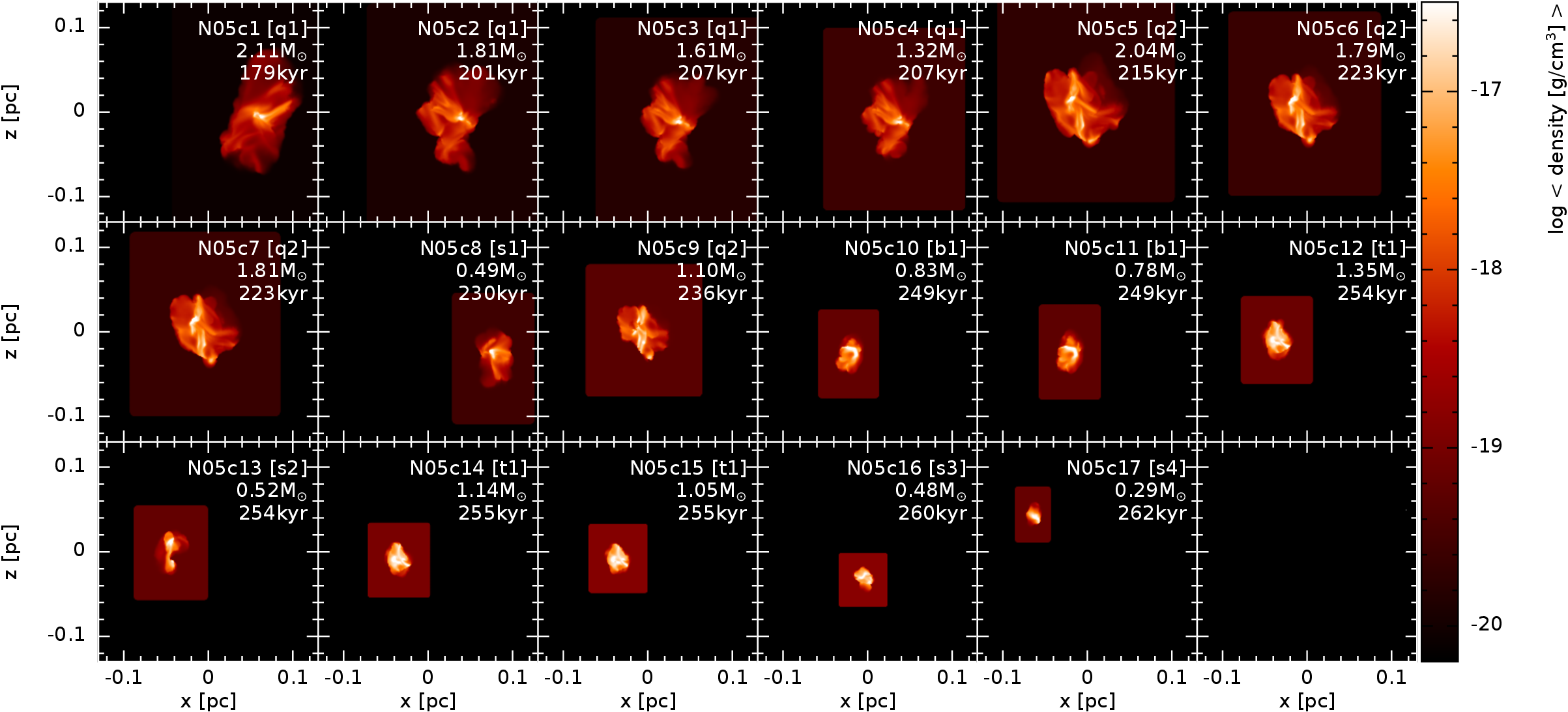}
\includegraphics[width=\textwidth]{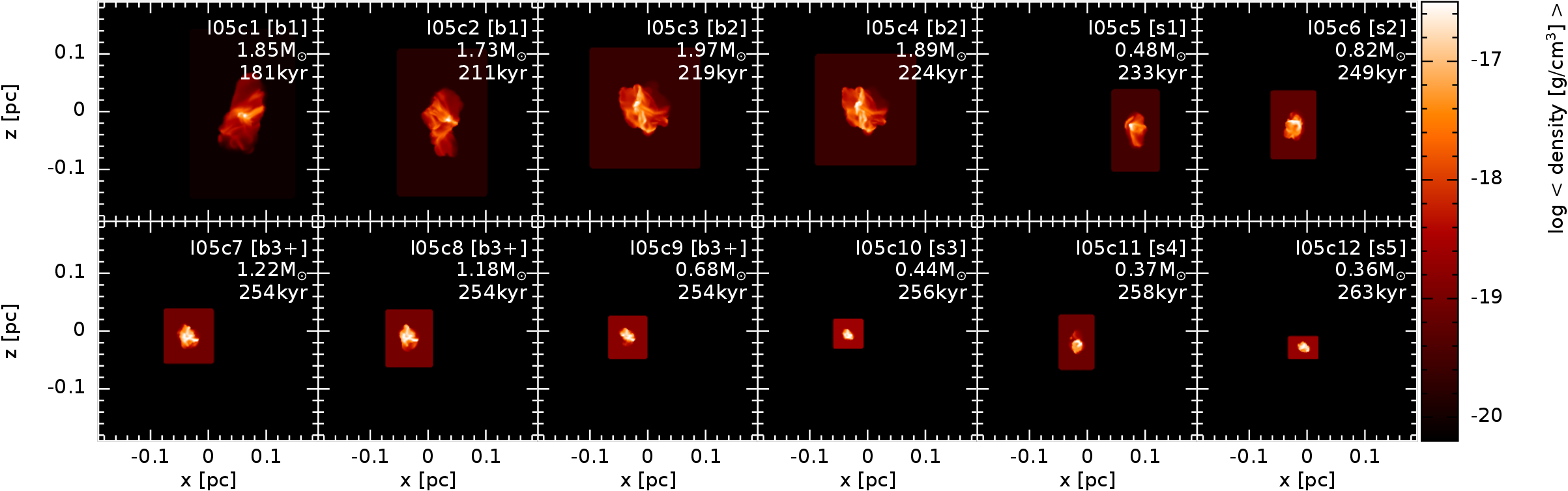}
\caption{Average gas density of clumps extracted from \nfive{} (top) and \ifive{} (bottom).  Each panel includes the clump name and name of the progenitor system, the mass of the clump, and the extraction time (as per \citetalias{WursterBatePrice2019}).  The clumps have not been re-centred, thus are a true indication of their extraction location in the parent cloud.  Clump mass typically decreases with increasing extraction time, even for stars that originated in the same progenitor system at \tfinal{}.  Clouds from the same progenitor system are qualitatively similar, but also generally vary in mass and extraction time. }
\label{fig:ics}
\end{figure*} 

Tables~\ref{tab:models:n} and \ref{tab:models:i} give mass of the gas common to the clumps extracted from the same progenitor systems in \nfive{} and \ifive{}, respectively.  The table with \sifive{b3+} includes the star that merged with \cifive{8} and the table with \sifive{s2,s4} compares the two spatially close but unbound stars in \ifive{}. In most systems, the smaller clumps are subsets of the most massive clump.  Exceptions are \snfive{q1} where \cnfive{1} is somewhat isolated, and \sifive{b1} where the two stars are independent despite being bound.

\begin{table}
\begin{center}
 \snfive{q1}\\
\begin{tabular}{c | c c c c c}
\hline
                &    \cnfive{1}        & \cnfive{2}       & \cnfive{3}       & \cnfive{4}       \\
\hline 
\cnfive{1}  & 2.11                  & 0.55               &    0.49             &    0.42          \\  
\cnfive{2}  &                         & 1.81               &      1.60             &    1.32         \\  
\cnfive{3}  &                         &                       &      1.61             &    1.31          \\  
\cnfive{4}  &                         &                       &                          &     1.32         \\  
\hline
\end{tabular}\\
\bigskip
 \snfive{q2}\\
\begin{tabular}{c c c c c c}
\hline
                 &    \cnfive{5}        & \cnfive{6}       & \cnfive{7}       & \cnfive{9}       \\
\hline 
\cnfive{5}  & 2.04                  & 1.78              &    1.80            &  1.10          \\ 
\cnfive{6}  &                          & 1.79              &    1.78            &  1.09         \\ 
\cnfive{7}  &                          &                       &   1.81            &  1.09          \\ 
\cnfive{9}  &                          &                      &                       &  1.10          \\
\hline
\end{tabular}\\
\bigskip
\snfive{t1}\\
\begin{tabular}{c c c c c }
\hline
                   &    \cnfive{12}        & \cnfive{14}       & \cnfive{15}          \\
\hline 
\cnfive{12}  & 1.35                   &  1.14               & 1.05        \\ 
\cnfive{14}  &                           &  1.14             &  1.04        \\
\cnfive{15}  &                           &                       &  1.05        \\
\hline
\end{tabular} \\
\bigskip
\snfive{b1}\\
\begin{tabular}{c c c c c }
\hline
                   &    \cnfive{10}        & \cnfive{11}        \\
\hline 
\cnfive{10}  & 0.83                 & 0.75                  \\ 
\cnfive{11}  &                         & 0.78                  \\
\hline
\end{tabular}
\caption{Each table presents the mass of the gas (in units of \Msun{}) common to the clumps extracted from the progenitor systems in \nfive{}.  The clump mass is given on the diagonal; the matrix is symmetric, so only the upper diagonal is completed for clarity.  Although \snfive{q1} is a quadruple system,  \cnfive{1} has a small fraction of gas in common with  \cnfive{2-4} thus is relatively isolated;   \cnfive{3-4} are sub-sets of \cnfive{2-3}.  Each clump in \snfive{q2} is a subset of each more-massive, older clump, and all clumps are a subset of \cnfive{5}.  Each clump in \snfive{t1} is a subset of each more-massive, older clump, and all clumps are a subset of \cnfive{12}.}
\label{tab:models:n} 
\end{center}
\end{table}

\begin{table}
\begin{center}
\sifive{b1} \\
\begin{tabular}{c c c c c }
\hline
                 &    \cifive{1}        & \cifive{2}        \\
\hline 
\cifive{1}  & 1.85                & 0.29                  \\ 
\cifive{2}  &                        &  1.73                    \\ 
\hline
\end{tabular}\\
\bigskip
\sifive{b2} \\
\begin{tabular}{c c c c c }
\hline
                &    \cifive{3}        & \cifive{4}        \\
\hline 
\cifive{3}  & 1.97                 & 1.86                  \\ 
\cifive{4}  &                         & 1.89                   \\
\hline
\end{tabular}\\
\bigskip
\sifive{b3+} \\
\begin{tabular}{c c c c c }
\hline
               &    \cifive{7}        & \cifive{8}       & \cifive{9}          \\
\hline 
\cifive{7} &1.22                    & 1.10               &   0.68        \\  
\cifive{8} &                           & 1.18               &  0.67        \\
\cifive{9}  &                          &                       &  0.68         \\
\hline
\end{tabular}\\
\bigskip
\sifive{s2,s4} \\
\begin{tabular}{c c c c c }
\hline
                 &    \cifive{6}        & \cifive{11}        \\
\hline 
\cifive{6}  & 0.82                  &  0.35                  \\ 
\cifive{11}  &                        &  0.37                  \\ 
\hline
\end{tabular}
\caption{Mass of the gas (in units of \Msun{}) common to the clumps extracted from the stars in \ifive{}.  The clump mass is given on the diagonal; the matrix is symmetric, so only the upper diagonal is completed for clarity.  Despite the stars in \sifive{b1} being bound, only \sm16 per cent of the gas is common between the clumps, therefore they are mostly independent.  \cifive{9} merged with \cifive{8}; nearly all of the gas in \cifive{9} is in the other two clumps, therefore, \cifive{7,8} also represent the merged star. Although  \cifive{6}  and \cifive{11} are not bound, nearly all the gas in \cifive{11} is also in \cifive{6}, therefore, the latter should well represent both progenitors.}
\label{tab:models:i} 
\end{center}
\end{table}


\label{lastpage}
\end{document}